\documentclass[prx,aps,twocolumn,showpacs,floatfix]{revtex4-1}
\usepackage{amsfonts}
\usepackage{amssymb}
\usepackage{hyperref}
\usepackage{graphicx}
\usepackage{dcolumn}
\usepackage{bm,amsmath,verbatim}
\usepackage{mathrsfs}
\usepackage{color}
\usepackage{lineno}
\usepackage{dsfont}
\usepackage{amsmath,amssymb,amsthm}

\usepackage[T1]{fontenc}
\usepackage[latin9]{inputenc}
\usepackage{booktabs}

\renewcommand\toprule{\hline\hline}
\renewcommand\bottomrule{\hline\hline}

\hypersetup{colorlinks,
linkcolor=blue,          citecolor=blue,        filecolor=blue,      urlcolor=blue           }

\def\be{\begin{equation}}
\def\ee{\end{equation}}
\def\bea{\begin{eqnarray}}
\def\eea{\end{eqnarray}}
\def\bse{\begin{subequations}}
\def\ese{\end{subequations}}

\def\be{\begin{eqnarray}}
\def\ee{\end{eqnarray}}

\begin{document}

%\setlength\columnsep{25pt}
%\linenumbers
%\def\linenumberfont{\tiny}

\title{Observation of topological Euler insulators with a trapped-ion quantum simulator}
\author{W.-D. Zhao}
\thanks{These authors contribute equally to this work}
\affiliation{Center for Quantum Information, Institute for Interdisciplinary Information Sciences, Tsinghua University, Beijing, 100084, PR China}

\author{Y.-B. Yang}
\thanks{These authors contribute equally to this work}
\affiliation{Center for Quantum Information, Institute for Interdisciplinary Information Sciences, Tsinghua University, Beijing, 100084, PR China}

\author{Y. Jiang}
\thanks{These authors contribute equally to this work}
\affiliation{Center for Quantum Information, Institute for Interdisciplinary Information Sciences, Tsinghua University, Beijing, 100084, PR China}

\author{Z.-C. Mao}
\affiliation{Center for Quantum Information, Institute for Interdisciplinary Information Sciences, Tsinghua University, Beijing, 100084, PR China}

\author{W.-X. Guo}
\affiliation{Center for Quantum Information, Institute for Interdisciplinary Information Sciences, Tsinghua University, Beijing, 100084, PR China}

\author{L.-Y. Qiu}
\affiliation{Center for Quantum Information, Institute for Interdisciplinary Information Sciences, Tsinghua University, Beijing, 100084, PR China}

\author{G.-X. Wang}
\affiliation{Center for Quantum Information, Institute for Interdisciplinary Information Sciences, Tsinghua University, Beijing, 100084, PR China}

\author{L. Yao}
\affiliation{Center for Quantum Information, Institute for Interdisciplinary Information Sciences, Tsinghua University, Beijing, 100084, PR China}

\author{L. He}
\affiliation{Center for Quantum Information, Institute for Interdisciplinary Information Sciences, Tsinghua University, Beijing, 100084, PR China}

\author{Z.-C. Zhou}
\email{zichaozhou@mail.tsinghua.edu.cn}
\affiliation{Center for Quantum Information, Institute for Interdisciplinary Information Sciences, Tsinghua University, Beijing, 100084, PR China}
%\affiliation{Beijing Academy of Quantum Information Sciences, Beijing 100193, PR China}

\author{Y. Xu}
\email{yongxuphy@mail.tsinghua.edu.cn}
\affiliation{Center for Quantum Information, Institute for Interdisciplinary Information Sciences, Tsinghua University, Beijing, 100084, PR China}

\author{L.-M. Duan}
\email{lmduan@mail.tsinghua.edu.cn}
\affiliation{Center for Quantum Information, Institute for Interdisciplinary Information Sciences, Tsinghua University, Beijing, 100084, PR China}

\begin{abstract}
Symmetries play a crucial role in the classification of topological phases of matter.
Although recent studies have established a powerful framework to search for and classify topological phases based on symmetry indicators,
there exists a large class of fragile topology beyond the description. The Euler class characterizing the topology of two-dimensional real wave
functions is an archetypal fragile topology underlying some important properties, such as non-Abelian braiding of crossing nodes
and higher-order topology. However, as a minimum model of fragile topology, the two-dimensional topological Euler insulator
consisting of three bands remains a significant challenge to be implemented in experiments.
Here, we experimentally realize a three-band Hamiltonian to simulate a topological Euler insulator with a trapped-ion quantum simulator.
Through quantum state tomography, we successfully evaluate the Euler class, Wilson loop flow and entanglement spectra to show the topological
properties of the Hamiltonian. We also measure the Berry phases of the lowest energy band, illustrating the existence of four crossing points
protected by the Euler class.
The flexibility of the trapped-ion quantum simulator further allows us to probe dynamical topological features
including skyrmion-antiskyrmion pairs and Hopf links in momentum-time space from quench dynamics.
Our results show the advantage of quantum simulation technologies for studying exotic topological phases
and open a new avenue for investigating fragile topological phases in experiments.
\end{abstract}

\maketitle

\section{Introduction}

Topological phases have seen a rapid progress over the past two decades~\cite{Hasan2010RMP,Qi2011RMP,Chiu2016RMP,Wen2017RMP,Vishwanath2018RMP}.
In particular, the ten-fold classification based on the $K$-theory represents a
cornerstone in the description of topological phases~\cite{Schnyder2008PRB,Kitaev2009AIP,Ryu2010NJP}.
Besides the internal symmetries, crystalline symmetries greatly enrich the
classification of topological phases~\cite{Fu2011PRL,Hughes2011PRB,Slager2013NP,Sato2014PRB}. Remarkably, recent efforts have led to the
development of a powerful framework based on symmetry indicators to categorize topological
crystalline insulators~\cite{Po2017NC,Bernevig2017Nature,Slager2017PRX}. However, a large class of topological phases falls outside the description~\cite{Po2018PRL,Bradlyn2019PRB,Ortix2019PRB,Bouhon2019PRB,BJYang2019PRX,Song2020PRX,Slager2020PRB}.
Such a phase belongs to the category of so called fragile topology that can be trivialized
by adding trivial bands~\cite{Po2018PRL}, in stark contrast to a stable topology which remains nontrivial upon adding trivial bands.
In this context, a class of topological phases protected by space-time inversion symmetry
is highlighted~\cite{YXZhao2017PRL,BJYang2018PRL,BJYang2019PRX,Wu2019Science,Slager2020NP,Slager2020PRB,YXZhao2020PRL,Chan2020Nature,Jiang2021NP}. Among them, the Euler class characterizing the topological property of
two-dimensional real wave functions
underlies the failure of the Nielsen-Ninomiya theorem~\cite{BJYang2019PRX} and the existence of Wilson loop winding~\cite{Slager2020PRB}. Yet, adding a
trivial band can annihilate crossing nodes through braiding and remove the Wilson loop winding,
showing a fragile topology of the Euler class. Such a fragile topology is theoretically shown to protect the nonzero
superfluid weight in twisted bilayer graphene~\cite{Xie2020PRL}.
Despite recent important progress on experimental characterizations of the fragile topology in an acoustic metamaterial~\cite{Huber2020Science},
the implementation of the topological Euler insulator~\cite{Slager2020PRL,Ezawa2021PRB} as a minimum model of fragile topology poses a significant experimental challenge.

Quantum simulators have been proven to be powerful platforms to experimentally study novel topological phases.
During the past decade, there have been great advances in simulating various topological phases via different quantum simulators
including cold atom systems~\cite{Esslinger2014Nature,Shuai2016Science,Spielman2018Science,Jo2018SA,Browaeys2019Science}, solid-state spin systems~\cite{Yuan2017CPL,2019Machine,Jiangfeng2020PRL,Dawei2020PRL,Wengang2021PRL} and superconducting circuits~\cite{Norman2017PRX,Luyan2019PRL,Shiliang2021PRL,Dapeng2021SciBull}.
Trapped ions provide an alternative flexible platform to perform quantum simulations due to its state-of-the-art
technologies to control and measure~\cite{Blatt2012NP,MonroeReview},
enabling us to use it to simulate
exotic topological phases
and directly probe their intriguing topological properties through measurements with high precision.

\begin{figure*}[t]
\centering
\includegraphics[width=1.0\textwidth]{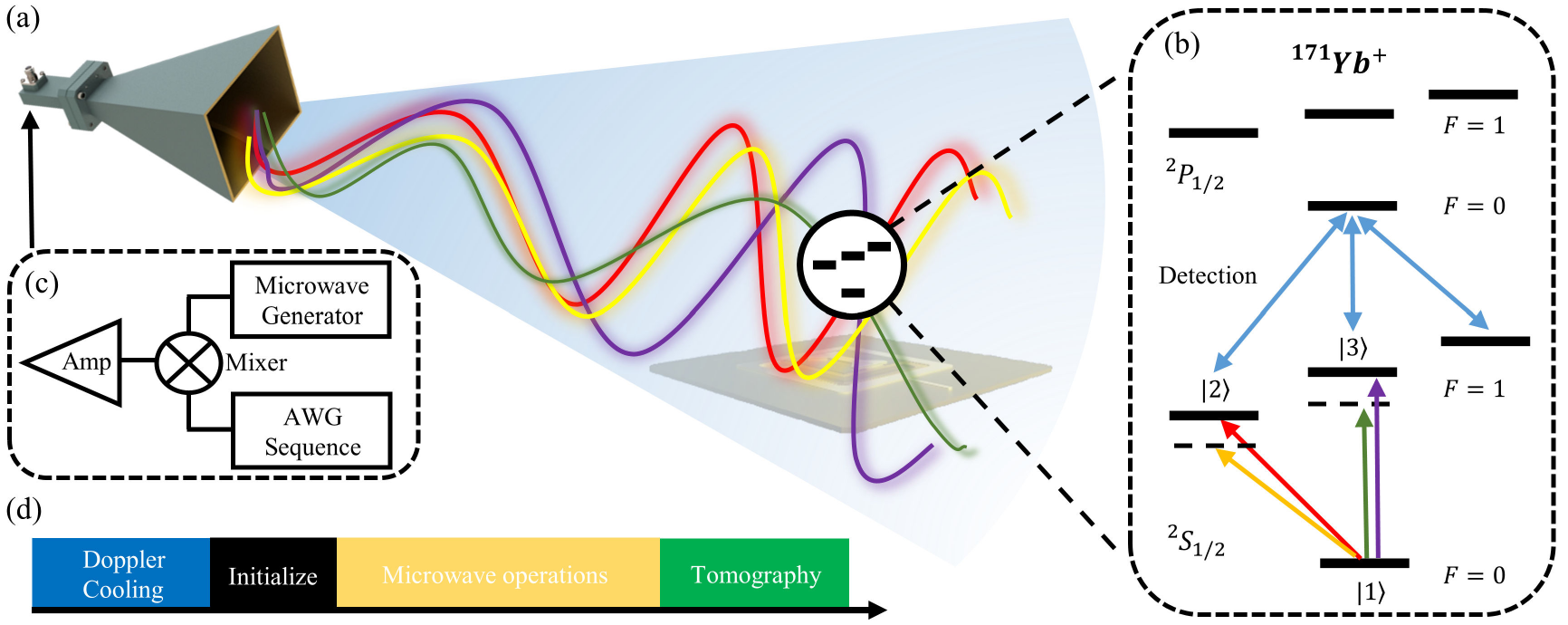}
\caption{
\textbf{Experimental scheme for observing topological Euler insulators with a trapped ion.}
(a) Schematic of our experimental setup. A single ${}^{171}$Yb$^{+}$ ion is trapped in a linear Paul trap realized by a surface-electrode chip trap. (b) The energy level structure of the ${}^{171}$Yb$^{+}$ ion. The three states $|1\rangle $, $|2\rangle $, and $|3\rangle $ for the Hamiltonian are encoded in the hyperfine states $|F=0,m_F=0\rangle $, $|F=1,m_F=-1\rangle $, and $|F=1,m_F=0\rangle $, respectively. Quantum operations and adiabatic evolutions are implemented by microwaves. Two far-detuned microwave pulses (denoted by yellow and green arrows) are used for Raman transitions between the levels $|2\rangle $ and $|3\rangle $. The quantum state projected to $|2\rangle $ or $|3\rangle $ generates fluorescence detected by a $370\,$nm detection beam. (c) Schematic of our microwave setup. The microwaves are generated by an arbitrary waveform generator (AWG) controlled by a PC according to the sequence mixed with a high frequency microwave signal. They are then amplified by a microwave horn and shone on the ion.  (d) Experimental sequences. The ion is first cooled and initialized to an eigenstate at a high-symmetry point in momentum space. We then follow the adiabatic passage to slowly tune the Hamiltonian by microwave operations so as to drive the state to an eigenstate of $H(\bm{k})$ at any momentum point in the 2D Brillouin zone. At the end, we perform the quantum state tomography to obtain the full density matrix of the final state.
}
\label{fig1}
\end{figure*}

In the article, we experimentally implement a three-band topological Euler Hamiltonian in momentum space
using a single ${}^{171}$Yb$^{+}$ ion trapped by a electrode-surface trap as shown in Fig.~\ref{fig1}.
By measuring the momentum-resolved eigenstates through quantum state tomography,
we evaluate the Euler class $\xi$, Wilson loop flow, and entanglement spectra
to identify the band topology.
With the Euler class, there are $2\xi$ protected crossing nodes between the two lowest energy bands.
Such nodes can be annihilated either by closing the gap between the second and third bands or by adding a trivial band,
followed by
intricate braiding of crossing nodes~\cite{BJYang2019PRX,Slager2020NP,Jiang2021NP}.
Our further measurements of the Berry phases along four distinct closed trajectories
illustrate the existence of four crossing points. Apart from the equilibrium topological properties,
it has been theoretically demonstrated that
skyrmion-antiskyrmion pairs and Hopf links appear in momentum-time space from quench dynamics
for the post-quench three-band topological Euler Hamiltonian~\cite{Slager2020PRL}.
We experimentally observe the skyrmion-antiskyrmion pairs and Hopf links by measuring the time-evolving states
under the Euler Hamiltonian.

The paper is organized as follows.
In Sec.~\ref{sec2}, we introduce our model Hamiltonian for topological Euler insulators
and demonstrate the experimental realization of the three-band Bloch Hamiltonian in a trapped-ion qutrit system.
In Sec.~\ref{sec3}, we present the measurement results of the Euler class as the topological invariant for the Euler Hamiltonian through quantum state tomography.
In Sec.~\ref{sec4}, we evaluate the Wilson loop and entanglement spectra to characterize the band topology of the Euler Hamiltonian.
In Sec.~\ref{sec5}, we show the existence of topologically protected band nodes between the lower two bands by measuring Berry phases along corresponding closed paths.
In Sec.~\ref{sec6}, we report the observation of dynamical topological structures from the time-evolving state measured during the quench dynamics under the Euler Hamiltonian.
Finally, the conclusion is presented in Sec.~\ref{sec7}.

\section{Euler Hamiltonian and its experimental realization}
\label{sec2}

We start by considering the following three-band Hamiltonian for Euler insulators
in momentum space, which will be experimentally engineered,
\begin{equation}\label{H1}
H(\bm{k})=2\bm{n}(\bm{k})\cdot\bm{n}(\bm{k})^{T}-I,
\end{equation}
where
\begin{align}\label{nk}
\bm{n}(\bm{k})
&=\left(n_x(\bm{k}),n_y(\bm{k}),n_z(\bm{k})\right)^T \nonumber \\
&=\frac{1}{\mathcal{N}}\left( m-\cos(k_x)-\cos(k_y),\sin(k_x),\sin(k_y) \right)^T
\end{align}
is a real unit vector at $\bm{k}=(k_x,k_y)$ in the 2D Brillouin zone.
Here, $\mathcal{N}$ is the normalization factor, and $m$ is a parameter with $|m|\neq 0,2$ for $H(\bm{k})$ to be well-defined.
The Hamiltonian is a real symmetric matrix and thus respects $C_2 \mathcal{T}$ (composition of twofold rotational and time-reversal operators) symmetry, which can be represented by the complex conjugation $\mathcal{K}$ in a suitable basis~\cite{Slager2020NP}.
For simplicity, we have flattened the spectrum of the Hamiltonian without affecting the band topology.
The Hamiltonian has two degenerate bands $ |u_{1,2}(\bm{k}) \rangle= \bm{u}_{1,2}(\bm{k})  $ with eigenenergy $E_{1,2}=-1$
and one band $|u_3(\bm{k})\rangle =\bm{n}(\bm{k})= \bm{u}_{1}(\bm{k}) \times \bm{u}_{2}(\bm{k})$ with eigenenergy $E_3=1$. The eigenstates are real unit vectors because of reality of the Hamiltonian.

We implement the Euler Hamiltonian $H(\bm{k})$ in momentum space through microwave operations on three hyperfine states
$|1\rangle=|F=0,m_F =0\rangle $, $|2\rangle=|F=1,m_F=-1\rangle $, and $|3\rangle=|F=1,m_F=0\rangle $ in the ${}^{2}S_{1/2}$ manifold
using a single ${}^{171}$Yb$^{+}$ ion trapped in an electrode-surface chip
trap as shown in Fig.~\ref{fig1} (see Appendix A for details).
In the experiment, a magnetic field is applied to the system to split the $|2\rangle$ and $|3\rangle$
levels so that we can individually control the couplings between the hyperfine states through microwave operations.
To control the amplitude and phase of microwaves, we use an arbitrary waveform generator (AWG) mixed with a high-frequency signal to modulate them.
Specifically, we drive
the transition between the $|1\rangle$ and $|2\rangle$ levels or the transition between the $|1\rangle$ and $|3\rangle$ levels
by near resonant microwaves and drive the transition between the $|2\rangle$ and $|3\rangle$ levels by two far-detuned
microwaves through microwave Raman transitions (see Appendix B for details). In the experiment, we in fact implement the Hamiltonian
${H}_{\textrm{exp}} (\bm{k}) = c [ H(\bm{k}) - b {I_3} ]$ ($I_3$ is the $3\times 3$ identity matrix) which has
the same eigenstates as $H(\bm{k})$. Here, $c$ is a numerically computed maximum value for our accessible system parameters at each $\bm k$
so that the energy gap takes a maximum value.

To measure the band topology of ${H}_{\textrm{exp}} (\bm{k})$,
we first prepare the ion in the dark state $|1\rangle$,
which is the highest
energy eigenstate of the Hamiltonian $H(\bm{k})$ at high-symmetry points ${\bm k}^*$ in momentum space:
$\bm{k}^*=(k_x, k_y) \in \{(0,0), (0,\pi), (\pi,0), (\pi,\pi)\}$.
We then slowly vary the Hamiltonian to the final one $H_{\textrm{exp}}(\bm{k})$ through the shortest path
in the Brillouin zone from the starting point $\bm{k}^*$ to the final point $\bm{k}$.
Since $|1\rangle$ is
the highest energy eigenstate $|u_3(\bm{k}^*)\rangle$ of our initial Hamiltonian $H(\bm{k}^*)$,
the state can evolve to a state which is very close to the highest energy
eigenstate $|u_3(\bm{k})\rangle$ of ${H}_{\textrm{exp}} (\bm{k})$ at the momentum ${\bm k}$ after the adiabatic passage.
At the end of microwave operations, we employ quantum state tomography to obtain the full density matrix $\rho(\bm{k})$
of the qutrit system (see Appendix C for details).
The fidelity for the measured density matrix is calculated as
$F(\bm{k}) = \langle \psi(\bm{k}) | \rho(\bm{k}) | \psi(\bm{k}) \rangle$,
where $\psi(\bm{k})$ is the theoretically obtained state, and $\rho(\bm{k})$ is the measured density matrix after optimization.
For the adiabatic preparation, the average fidelity is about $97\%$.
To identify the band topology of the Euler Hamiltonian, we
need to transform the measured density matrix into a real state that is closest to $\rho(\bm{k})$
as the measured state for $|u_3(\bm{k})\rangle$ (see Appendix D for details).
Since $|u_3(\bm{k})\rangle$ contains full information of the flattened Hamiltonian $H(\bm{k})$, it enables us to determine the topological properties of the Euler Hamiltonian using these measured states.

\section{Euler class}
\label{sec3}

\begin{figure}[t]
\centering
\includegraphics[width=1.0\linewidth]{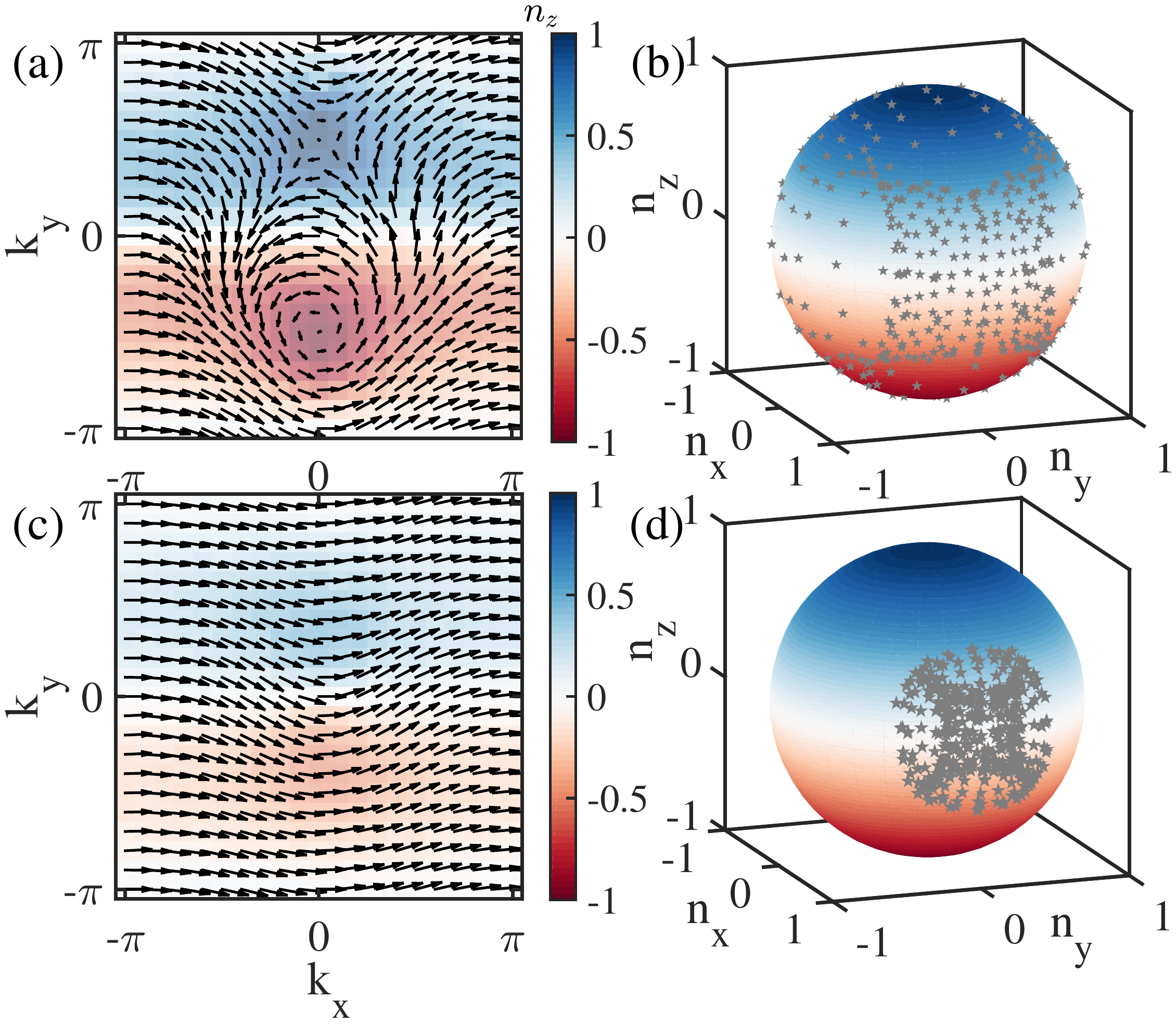}
\caption{\textbf{Measurements of the Euler class.}
(a,c) The experimentally measured vectors $\bm{n}(\bm{k})$ for the Euler Hamiltonian $H(\bm{k})$ (\ref{H1}) over a $20\times 20$ discretized Brillouin zone and
(b,d) their distribution on the sphere $S^2$. In (a,b), $m=1$ corresponding to a nontrivial Hamiltonian and
in (c,d), $m=3$ corresponding to a trivial one.
In the nontrivial case, the vectors $\bm{n}(\bm{k})$ exhibit a nontrivial skyrmion structure and
cover the entire sphere once, yielding a nonzero Euler class $\xi=2$, whereas in the trivial case,
the vectors cover parts of the sphere, yielding a zero Euler class.
}
\label{fig2}
\end{figure}

The band topology of a Euler insulator can be characterized by the Euler class~\cite{YXZhao2017PRL,BJYang2018PRL,BJYang2019PRX,Slager2020NP}
\begin{equation}
\xi=\frac{1}{2\pi}\int_{BZ} \mathrm{d}^2 \bm{k}
(\langle \partial_{k_x} u_1 | \partial_{k_y} u_2 \rangle -
\langle \partial_{k_y} u_1 | \partial_{k_x} u_2 \rangle)
\end{equation}
for the two occupied bands $|u_1\rangle$ and $|u_2\rangle$.
It is enforced to be quantized by the reality of eigenstates due to $C_2 \mathcal{T}$ symmetry.
To well define the Euler class, we require that the two occupied bands form an orientable real vector bundle,
which is ensured by the vanishing of Berry phases along any noncontractible loops for the occupied space~\cite{BJYang2019PRX,Slager2020NP,Slager2020PRB}.
For a three-band Euler Hamiltonian,
the Euler class can be reduced to another form~\cite{Slager2020NP}
\begin{equation}\label{Euler}
\xi=\frac{1}{2\pi}\int_{BZ} \mathrm{d}^2 \bm{k} [\bm{n}\cdot(\partial_{k_x}\bm{n}\times \partial_{k_y}\bm{n})],
\end{equation}
showing that $\xi$ is equal to twice of the winding number (Pontryagin number) of $\bm{n}(\bm{k})$ over the 2D Brillouin zone
(see Appendix E). In other words,
the Euler class is determined by twice of times that $\bm{n}(\bm{k})$ wraps the sphere $S^2$,
which also implies that the Euler class for a three-band Hamiltonian should be an even integer.
Note that if the orientations of all vectors $\bm{n}(\bm{k})$ are reversed, we
obtain the same Hamiltonian but the opposite winding for $\bm{n}(\bm{k})$, showing that
the sign of $\xi$ is ambiguous and only its absolute value characterizes the topology~\cite{Slager2020PRB}.
With the relation between the Euler class $\xi$ and the winding number of $\bm{n}(\bm{k})$,
we can obtain the phase diagram that the Euler Hamiltonian $H(\bm{k})$ (\ref{H1})
is topologically nontrivial with $\xi=2$ for $0<|m|<2$ and trivial with $\xi=0$ for $|m|>2$.

Our experimentally measured vectors $\bm{n}(\bm{k})$ in a topological phase indeed exhibit a skyrmion structure
over the Brillouin zone [see Fig.~\ref{fig2}(a)] wrapping the entire sphere once [see Fig.~\ref{fig2}(b)],
which suggests that the Euler class $\xi=2$ for the experimentally realized Hamiltonian.
To be more quantitative, we map the measured vectors to
a two-band Chern insulator by $H_C(\bm{k})=\bm{n}(\bm{k}) \cdot \bm{\sigma}$
with Pauli matrices $\bm{\sigma}=(\sigma_x,\sigma_y,\sigma_z)$. The fact that the Euler class is
equal to twice of the Chern number of $H_C(\bm{k})$ allows us to determine the Euler
class by computing the Chern number, which is much more efficient than directly performing the integral for
Eq.~(\ref{Euler}) (see more details in Appendix E).
We find that the Chern number calculated using the measured $\bm{n}(\bm{k})$ is equal to $1$ so that $\xi=2$.
In comparison, we also display the measured vectors $\bm{n}(\bm{k})$ in a trivial phase,
which do not form a skyrmion structure [see Fig.~\ref{fig2}(c)]
and cover only parts of the sphere [see Fig.~\ref{fig2}(d)], indicating that $\xi=0$.

\section{Wilson loops and entanglement spectra}
\label{sec4}

\begin{figure}[t]
\centering
\includegraphics[width=3.3in]{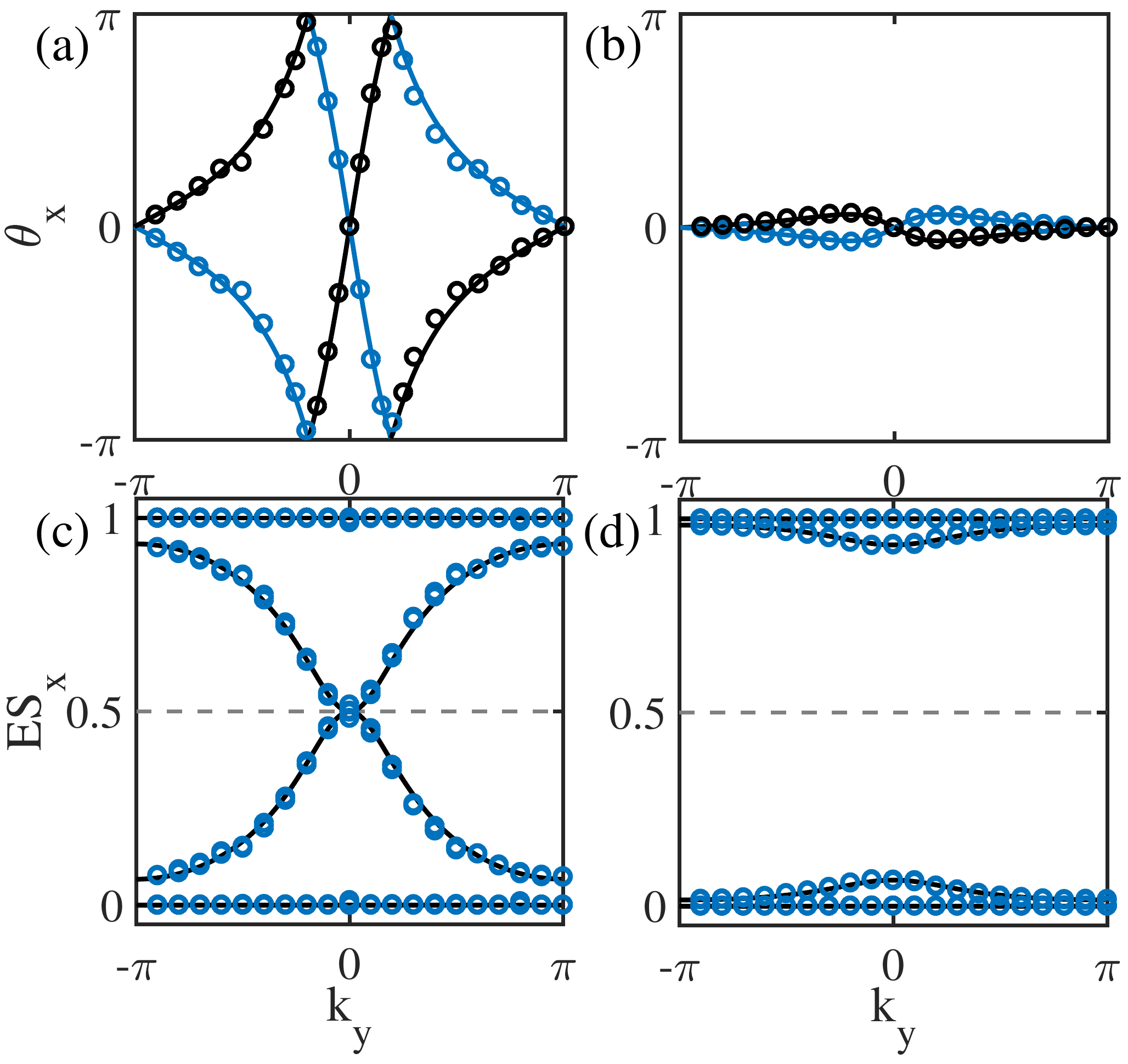}
\caption{\textbf{Observation of Wilson loop spectra and entanglement spectra.}
(a-b) The experimentally measured Wilson loop spectra (solid circles) in comparison with the numerically calculated ones (solid lines).
(c-d) The experimentally measured entanglement spectra (circles) in comparison with the numerical ones (solid lines).
We consider the nontrivial Euler Hamiltonian $H(\bm{k})$ (\ref{H1}) with $m=1$ in (a) and (c) and the trivial one
with $m=3$ in (b) and (d).
}
\label{fig3}
\end{figure}

The Wilson loop provides a powerful framework to characterize the fragile topology~\cite{BJYang2019PRX,Bradlyn2019PRB,Bouhon2019PRB,Slager2020PRB}.
However, it is very challenging to
make an experimental measurement of it. In the trapped-ion quantum simulator, the quantum state tomography technique allows us to
evaluate the Wilson loop based on the measured eigenstates of a Euler Hamiltonian.

Specifically, the $x$-directed Wilson loop $\mathcal{W}_x$ with the base point $\bm{k}_0=(k_x,k_y)$
can be computed in a $N \times N$ discretized Brillouin zone as~\cite{Bernevig2014PRB}
\begin{equation}
[\mathcal{W}_x]_{mn}=\langle u_m(\bm{k}_0) | \prod_{\bm{k}_j}^{\bm{k}_N \leftarrow \bm{k}_1}
P_{\textrm{occ}}(\bm{k}_j) | u_n(\bm{k}_0) \rangle,
\end{equation}
where $\bm{k}_j=(k_x+\frac{2\pi j}{N},k_y)$, $N$ is the number of discrete momenta on the loop,
and $P_{\textrm{occ}}(\bm{k}_j)$ is the projector on the occupied bands at the momentum $\bm{k}_j$.
The Wilson loop operator $\mathcal{W}_x$ is unitary so that its eigenvalues take the form of $e^{i\theta_x(k_y)}$ which only depends on $k_y$;
$\theta_x(k_y)$ as a function of $k_y$ is known as the Wilson loop spectrum for $\mathcal{W}_x$.
The $y$-directed Wilson loop $\mathcal{W}_y$ and the corresponding spectrum $\theta_y(k_x)$ can be defined similarly.
For our three-band model, we only need the experimentally measured highest energy eigenstates to construct
the occupied projectors in the Wilson loop,
\begin{equation}
 P_{\textrm{occ}}(\bm{k}_j)
= I-|u_3(\bm{k}_j)\rangle\langle u_3(\bm{k}_j)|.
\end{equation}
The Wilson loop spectrum can be determined by diagonalizing the matrix
\begin{equation}
P= \prod_{\bm{k}_j}^{\bm{k}_N \leftarrow \bm{k}_0}
P_{\textrm{occ}}(\bm{k}_j),
\end{equation}
which gives us three eigenvalues as $\{e^{i\theta_x^{(1)}}, e^{i\theta_x^{(2)}},0 \}$.
Discarding the zero eigenvalue contributed by the unoccupied subspace,
we obtain the Wilson loop eigenvalues $\{\theta_x^{(1)},\theta_x^{(2)} \}$ for the two occupied bands.

For the Euler Hamiltonian with two occupied bands,
due to the reality of eigenstates, the Wilson loop operator takes the form of $e^{i\theta\sigma_y}$,
which has a pair of eigenvalues $e^{\pm i \theta}$.
For a topological Euler insulator, both $\theta_x(k_y)$ and $\theta_y(k_x)$ exhibit a nontrivial winding,
indicating an obstruction to the Wannier representation.
The winding number is equivalent to the Euler class $\xi$~\cite{Slager2020PRB}.

Figure~\ref{fig3}(a-b) shows the experimentally measured Wilson loop spectra $\theta_x(k_y)$ [$\theta_y(k_x)$ has similar behaviours],
which are evaluated based on the measured highest energy eigenstates.
In the topological phase, each branch of the Wilson
loop spectra exhibits a winding number ($\pm 2$),
whereas in the trivial phase, the winding patterns are not observed.
All the experimental results are in excellent agreement with the theoretical ones.
We remark that such a winding can be removed by adding a trivial band, which reveals the fragile
topology feature of the system (see Appendix F for more details).

Although we experimentally realize the topological Euler Hamiltonian in momentum space,
we can extract the edge state information through the single-particle entanglement spectra evaluated based on the measured
states; such spectra can exhibit more robust nontrivial features than those for physical boundaries in a topological band insulator~\cite{Fidkowski2010PRL,Turner2010PRB,Hughes2011PRB}.

Figure~\ref{fig3}(c) displays the entanglement spectra $\mathrm{ES}_x(k_y)$
obtained by partially tracing out the right part of the system
using the experimentally measured unoccupied eigenstates $|u_3(\bm{k})\rangle$ for the Euler Hamiltonian $H(\bm{k})$
(see Appendix G for more details).
In the topological phase, an in-gap spectrum with mid-gap modes near $\xi_n=0.5$ arises in the entanglement spectra,
which agrees very well with the theoretical prediction.
The experimental results also support the theoretical prediction of the parabolic dispersion for the entanglement spectra
near the mid-gap modes.
In the trivial phase, our experimental results do not reveal the existence of gapless entanglement spectra,
indicating that the phase is adiabatically connected to a trivial phase with zero entanglement entropy.

\section{Dirac points}
\label{sec5}

\begin{figure}[t]
\centering
\includegraphics[width=1.0\linewidth]{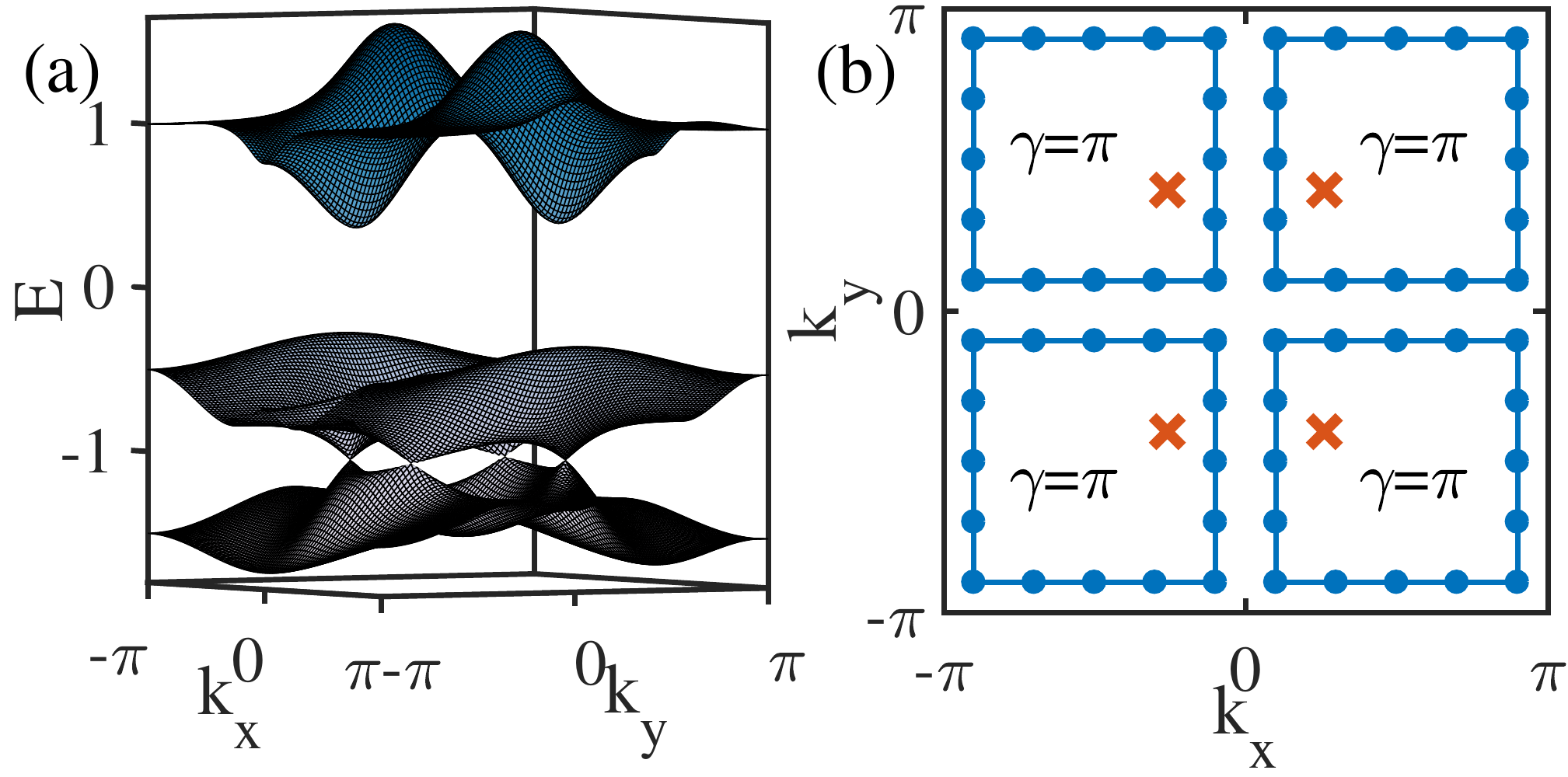}
\caption{\textbf{Measurements of the Berry phase.}
(a) The theoretical energy spectrum of the Hamiltonian $H'(\bm{k})$ (\ref{H2}) over the Brillouin zone,
showing the existence of four Dirac points between the two lowest energy bands located at $\bm{k}=(k_x,k_y)\approx(\pm 0.26\pi,\pm 0.4\pi)$.
(b) The experimentally measured Berry phases for the lowest energy eigenstate $|u_1(\bm{k})\rangle$
along four closed paths composed of discrete momenta (blue lines and points)
in the Brillouin zone enclosing the corresponding Dirac points (red crosses).
For each Dirac point, the measured Berry phase is $\gamma=\pi$.
}
\label{fig4}
\end{figure}

The Euler class $\xi$ is also manifested in the existence of $2\xi$ stable Dirac points
between the two occupied bands~\cite{BJYang2019PRX,Slager2020NP}. They are protected by the $C_2 \mathcal{T}$ symmetry
and cannot be annihilated without the gap closing with the third band.
To see this feature, we consider the following model by adding an extra term to $H(\bm{k})$ (\ref{H1}) with $m=1$,
\begin{equation}\label{H2}
H'(\bm{k})=H(\bm{k})+
\mathrm{diag}(h_0(\bm{k}),h_{+}(\bm{k}),h_{-}(\bm{k}))
\end{equation}
with $h_0(\bm{k})=0.1[\cos(k_y)-\cos(k_x)]$
and $h_{\pm}(\bm{k})=h_0(\bm{k})\pm 0.5$.
The additional term lifts the degeneracy of the two occupied bands for the flattened Hamiltonian $H(\bm{k})$
except at the Dirac points as shown in Fig.~\ref{fig4}(a).
Due to the $C_2 \mathcal{T}$ symmetry, a Dirac point between the two lowest bands yields a quantized Berry phase $\gamma=\pi$
for the lowest eigenstates $|u_1(\bm{k})\rangle$ along a closed path $l$
enclosing it.

Since the energy gap between the two lowest eigenstates on a path enclosing the Dirac point is opened,
we can still use the adiabatic passage to realize the eigenstate $|u_1(\bm{k})\rangle$ at the momenta on the closed path.
After that, we
measure the states by quantum state tomography
and then evaluate the Berry phase based on the measured states.
We find that the experimentally evaluated Berry phase $\gamma=\pi$ on the four closed paths [see Fig.~\ref{fig4}(b)],
indicating the presence of
a Dirac point inside each closed path.

\section{Quench dynamics}
\label{sec6}

Besides the equilibrium features, it has been theoretically shown that nonequilibrium dynamics provides another tool to uncover the static band topology of a Euler insulator~\cite{Slager2020PRL}.
Let us start from an initial state $ \psi(\bm{k},t=0) = \psi_0 =(0,0,1)^T$ for each momentum $\bm{k}$,
which can be seen as the eigenstate of a topologically trivial Euler Hamiltonian $H_0(\bm{k})=\mathrm{diag}(-1,-1,1)$.
We consider the quench dynamics for the trivial initial state evolving under a postquench Euler Hamiltonian $H(\bm{k})$ (\ref{H1}).
Due to the flatness of $H(\bm{k})$ (\ref{H1}),
the state evolves as
\begin{equation}
\psi(\bm{k},t) =[\cos(t)-i\sin(t)H(\bm{k})] \psi_0 ,
\end{equation}
which is periodic about time $t$ with a period $T=\pi$.
The periodicity of the evolving state both in time $t$ and momentum $\bm{k}$ in 2D Brillouin zone
makes the space of $(k_x,k_y,t)$ form a 3-torus $T^3$.
In analogy to the quench dynamics of a two-band Chern insulator~\cite{Zhai2017PRL},
one can construct a map $f$ from the $(k_x,k_y,t)$ space as a $T^3$ to a 2-sphere $S^2$ as follows.
For a point $(k_x,k_y,t)$ on $T^3$, the image of the map $f(k_x,k_y,t)$
is a unit vector $\hat{\bm{p}}=(p_x,p_y,p_z)$ on $S^2$ as
$\hat{\bm{p}}= \psi^{\dagger}(k_x,k_y,t)  \bm{\mu}  \psi(k_x,k_y,t) $,
where $\bm{\mu}=(\mu_x,\mu_y,\mu_z)$ with $\mu_\nu$ ($\nu=x,y,z$) being a $3\times3$ matrix~\cite{Slager2020PRL} (see Appendix H).
Because the map $f$ from any $T^2$ cross-section of $(k_x,k_y,t)$ space to $S^2$ is trivial with zero Chern number,
the map $f$ is equivalent to the form of a Hopf map from $S^3$ to $S^2$ classified by an integer called Hopf invariant~\cite{Moore2008PRL,Deng2013PRB}.
In the quench dynamics of a topological Euler Hamiltonian,
the Hopf invariant determines the linking number of a linking structure for the inverse images of the Hopf map~\cite{Slager2020PRL}.

\begin{figure}[t]
\centering
\includegraphics[width=1.0\linewidth]{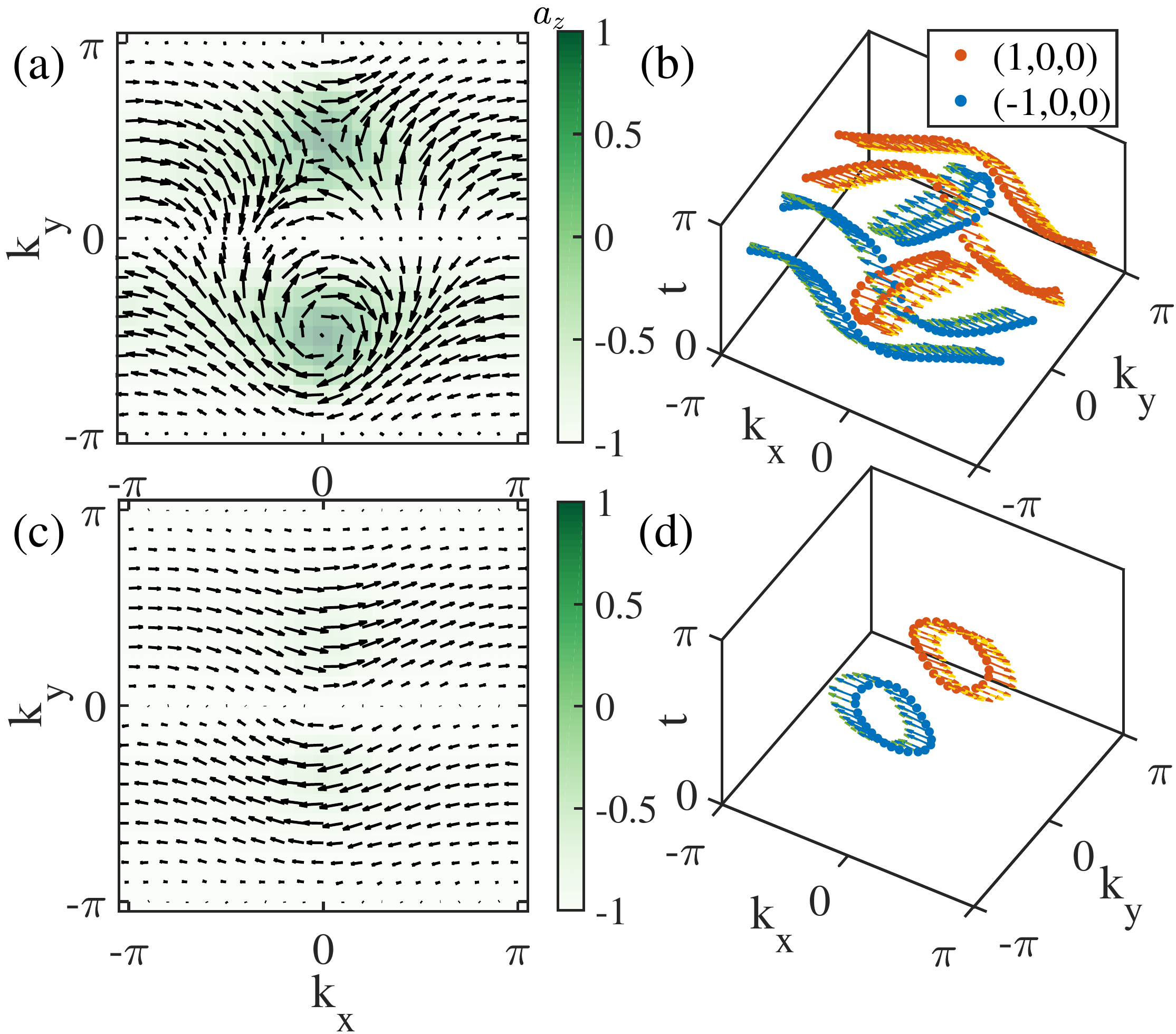}
\caption{
\textbf{Measurements of dynamical topological structures from quench dynamics.}
(a) The skyrmion-antiskyrmion structures for the vectors $\bm{a}(\bm{k})$ obtained by measuring the time-evolving state
at $t=\pi/2$ under the nontrivial postquench Hamiltonian $H(\bm{k})$ with $m=1$.
The skyrmion or antiskyrmion structures appear in each half of the 2D Brillouin zone,
contributing a Chern number of $\pm 1$, respectively.
(b) The skyrmion and antiskyrmion structure is directly associated with the nontrivial linking structure with a link and an antilink composed of
the inverse images $f^{-1}(\hat{\bm{p}}_1)$ and $f^{-1}(\hat{\bm{p}}_2)$ in $(k_x,k_y,t)$ space for
any two distinct points $\hat{\bm{p}}_1$ and $\hat{\bm{p}}_2$
on $S^2$. Here we take $\hat{\bm{p}}_1=(1,0,0)$ and $\hat{\bm{p}}_2=(-1,0,0)$.
The red and blue arrows show the images of the experimentally measured evolving state through the Hopf map,
which are close to the theoretical values marked by yellow and green arrows.
(c),(d) Quench dynamics for the trivial postquench Hamiltonian $H(\bm{k})$ with $m=3$.
The vectors $\bm{a}(\bm{k})$ have a topologically trivial distribution and the inverse images of $\hat{\bm{p}}_1$ and $\hat{\bm{p}}_2$ do not link with each other.
}
\label{fig5}
\end{figure}

The nontrivial linking structure for the quench dynamics of a Euler insulator is directly related to the static band topology of the postquench Hamiltonian~\cite{Slager2020PRL}. To see this relation, we write the evolving state as
\begin{equation}
\psi(\bm{k},t) =\cos(t) \psi_0  -i\sin(t) \bm{a}(\bm{k}).
\end{equation}
Here $\bm{a}(\bm{k})= H(\bm{k}) \psi_0 =i\psi(\bm{k},t=\pi/2)$, which defines a map from the 2D Brillouin zone to $S^2$.
Based on the static Hamiltonian $H(\bm{k})$ (\ref{H1}), we obtain
$\bm{a}(\bm{k})= (2n_x(\bm{k}) n_z(\bm{k}), 2n_y(\bm{k}) n_z(\bm{k}), 2n_z^2(\bm{k})-1)^T$.
By parameterizing $\bm{n}(\bm{k})$ with spherical coordinates
as $\bm{n}(\bm{k})=(\sin \alpha \cos \beta , \sin \alpha \cos \beta , \cos \alpha )$,
we have $\bm{a}(\bm{k})=(\sin 2\alpha \cos \beta , \sin 2\alpha \cos \beta , \cos 2\alpha )$.
For a nontrivial Euler Hamiltonian with $\xi=2$, $\bm{n}(\bm{k})$ fully cover the $S^2$
so that there exist 1D curves with $n_z(\bm{k})=\cos \alpha(\bm{k})=0$ in the Brillouin zone on which $\bm{a}(\bm{k})=(0,0,-1)$;
the curves divide the entire Brillouin zone into two patches.
The curves also serve as fixed points for the dynamics where the initial state only picks up a global phase during the time evolution.
By shrinking the curve into a single point, each of the two patches can be seen as a sphere.
In this case, $\bm{a}(\bm{k})$ defines a map from $S^2$ to $S^2$ characterized by the winding number for each of these patches.
Though the winding number of $\bm{a}(\bm{k})$ over the entire Brillouin zone is zero since the state $\psi(\bm{k},t=\pi/2)$ is trivial,
$\bm{a}(\bm{k})$ can wrap the sphere $S^2$ once in each patch~\cite{Slager2020PRL}.
The nontrivial winding of $\bm{a}(\bm{k})$ over each patch is associated with the Hopf link in the quench dynamics of the patch,
similar to the correspondence between the static Chern number and the existence of the dynamical Hopf link
for quench dynamics of a Chern insulator (see Appendix H).

Figure~\ref{fig5} shows our experimentally measured vectors ${\bm a}({\bm k})$ and linking structures.
Specifically, we first prepare the ion in the $|3\rangle$ level
and then measure the density matrix $\rho(\bm{k},t)$ of the time-evolving state
for a momentum in the Brillouin zone via quantum state tomography after the unitary time evolution
under the experimentally engineered Euler Hamiltonian.
We then evaluate $\bm{a}(\bm{k})$ and the images of the Hopf map, that is, $\langle \mu_i \rangle= \mathrm{Tr}(\rho(\bm{k},t) \mu_i)$ with $i=x,y,z$
based on the measured density matrices.
For the nontrivial postquench Euler Hamiltonian,
the experimentally measured $\bm{a}(\bm{k})$ exhibit a nontrivial skyrmion and antiskyrmion structure
in the upper and lower halves of the Brillouin zone divided by the curves $k_y=0,\pi$ with $n_z(\bm{k})=0$, as shown in Fig.~\ref{fig5}(a).
To quantitatively identify the skyrmion and antiskyrmion structure of the measured ${\bm a}({\bm k})$,
we construct a model $H_C(\bm{k})=\bm{a}(\bm{k})\cdot \bm{\sigma}$ for each of the two patches
and find that the Chern numbers are equal to $\pm 1$, which are in excellent agreement with the theoretical results.
The pair of skyrmions in the Brillouin zone leads to a pair of links with opposite signs
for the inverse images in the corresponding regions of the $(k_x,k_y,t)$ space (see Appendix H), which are experimentally demonstrated in Fig.~\ref{fig5}(b). This also indicates the nontrivial band topology
of the postquench Hamiltonian.
For a trivial postquench Hamiltonian, the measured $\bm{a}(\bm{k})$ do not fully cover $S^2$ for the two patches of the Brillouin zone,
and the inverse images have no linking structures, as shown in Fig.~\ref{fig5}(c) and (d).

\section{Conclusion}
\label{sec7}

We have experimentally realized a minimum Bloch Hamiltonian for topological Euler insulators protected by $C_2 \mathcal{T}$ symmetry
in a trapped-ion qutrit system
and identified its band topology by evaluating the Euler class, Wilson loop flow, entanglement spectra and
the Berry phases based on the measured states via quantum state tomography.
We further observed the nontrivial dynamical topological structures including the skyrmion-antiskyrmion structures and Hopf links
during the unitary evolution under the topological Euler Hamiltonian.
Our work opens the door for further studying fragile topological phases using quantum simulation technologies.

\begin{acknowledgments}
We thank W.-Q. Lian for helpful discussions. This work was supported by the Beijing Academy of Quantum Information Sciences, the Frontier Science Center for Quantum Information of the Ministry of Education of China, and Tsinghua University Initiative Scientific Research Program.
Y. Xu also acknowledges the support from the National Natural Science Foundation
of China (Grant No. 11974201).
\end{acknowledgments}

\section*{\label{sec:Experiment-Details}Appendix A: Experimental details}

\setcounter{equation}{0}
\renewcommand{\theequation}{A\arabic{equation}}
\renewcommand{\theHequation}{A\arabic{equation}}
\renewcommand{\bibnumfmt}[1]{[#1]} \renewcommand{\citenumfont}[1]{#1}

\begin{figure*}
  \centering
  \includegraphics[width=\textwidth]{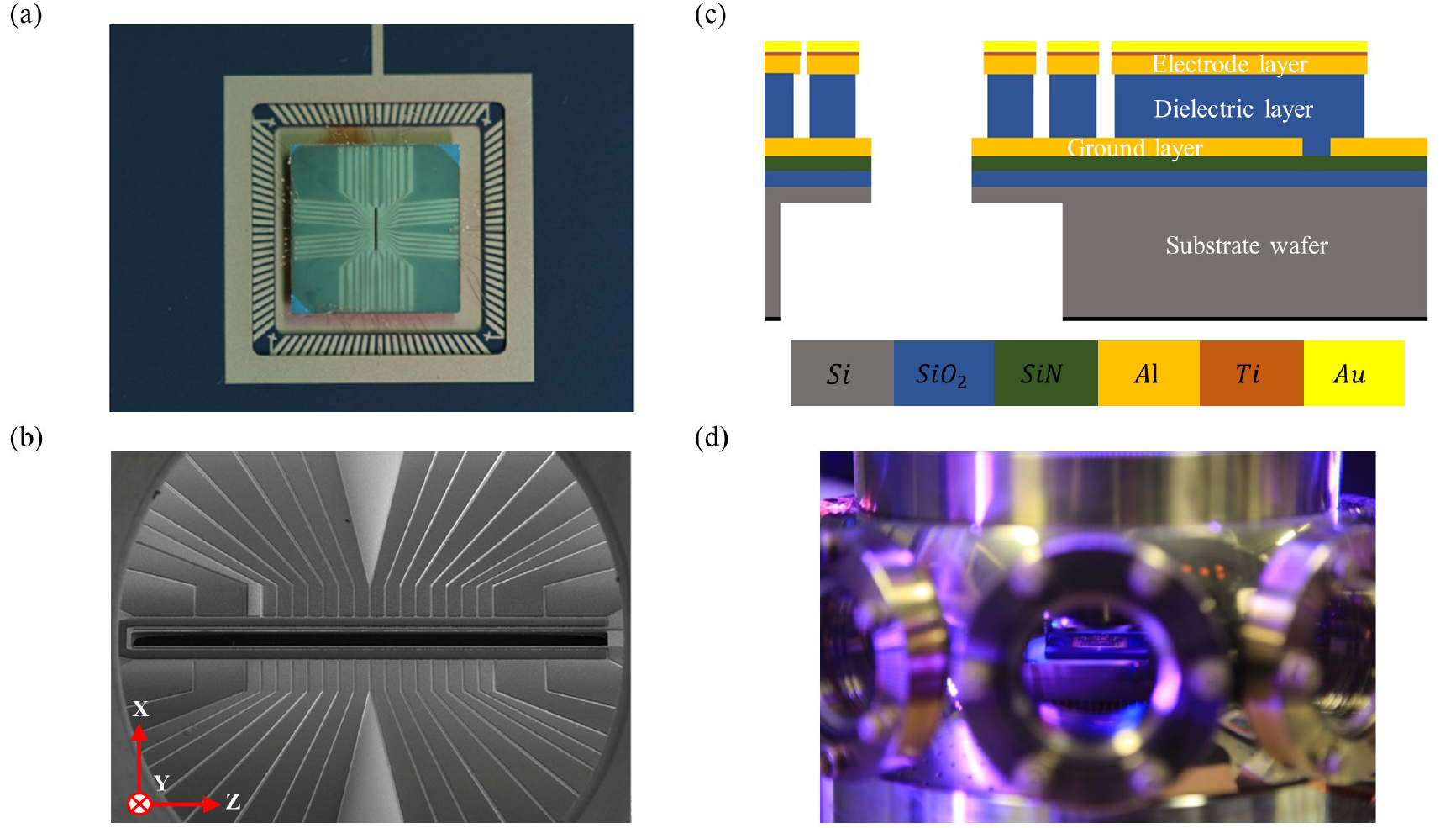}\\
  \caption{\textbf{Schematics of a surface-electrode trap chip.} (a) Top view of a surface trap fabricated in our group at
  CQI, IIIS, Tsinghua University. We use aluminum electrodes covered with gold on the top layer.
  The ion trap built on a silicon (Si) substrate is integrated into CPGA packaging
  by connecting the electrodes using bonding wires to bonding pads.
  (b) Scanning electron micrography of a surface trap chip. There is an axial loading slot in the center of the chip with
  its width being $102\mu$m, allowing for loading ions from the back side of the wafer and focusing individual
  lasers on ions directed perpendicular to the surface.
  We place a pair of global RF electrodes (with the width being $58\mu$m) with the RF electric signal on both sides of the slot
  to generate the Paul trap along the radial direction.
  In addition, a pair of
  global inner DC electrodes (with the width being $20\mu$m) and 20 pairs of outer segmented DC electrodes are attached to generate the axial confinement and shuttle ions. The length of the segmented DC electrodes along the axial direction is $62\mu$m.
  (c) A surface trap chip consists of a substrate wafer, a ground layer (wire guiding), a semiconductor dielectric layer, and an electrode layer from bottom to top.
  (d) An optical image of our vacuum chamber. }\label{fig:Chip_figure}
\end{figure*}

\subsection*{1. Micro-fabricated surface-electrode chip traps}

\begin{figure*}
\centering
\includegraphics[width=\textwidth]{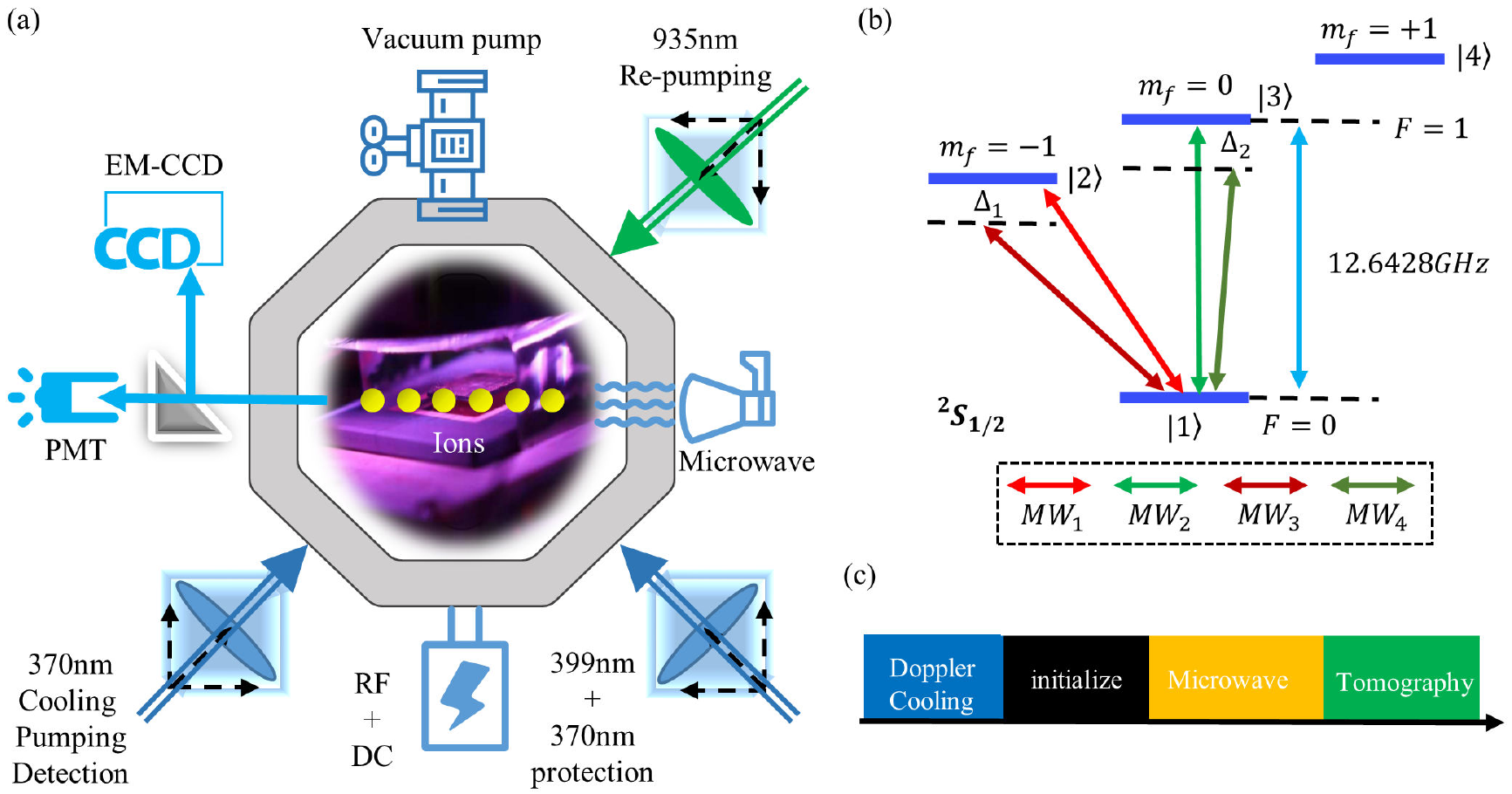}\\
\caption{\textbf{Schematics of the experimental system, energy level structures and experimental sequences.}
(a) Schematics of the experimental system. Ions are initially Doppler-cooled by $370\,$nm cooling beams to form a linear ion chain
along the $z$ axis or a single ion. Optical pumping beams and detection beams are combined with cooling beams, entering the octagon chamber from the left bottom window. The $399\,$nm and $935\,$nm laser beams are used for ionization and repumping. The fluorescence of ions is collected by an objective on the top of the chamber and transformed to electric signals by a PMT or EM-CCD. A RF signal is applied to the surface trap via a helical resonator for impedance matching. A horn on the right-hand side amplifies the microwaves, which are further shone to the ions.
(b) Energy level structures of the ${}^{171}$Yb$^{+}$ ion. Three states $|1\rangle$ , $|2\rangle$ and $|3\rangle$ are encoded in the $|F=0,m_F =0\rangle$, $|F=1,m_F=-1\rangle$ and $|F=1,m_F=0 \rangle$ in the ${}^{2}S_{1/2}$ manifold, respectively.
The energy splitting between three levels of $F=1$ states is determined by the strength of magnetic filed at the ion's position.
The transitions
between $F=0$ and $F=1$ states (whose central transition frequency is $\omega_{13} = 2\pi \times 12.6428335 \,$GHz) are driven by microwaves from a horn: two near resonant microwaves are used to realize the couplings between $|1\rangle$ and $|2\rangle$ or $|1\rangle$ and $|3\rangle$,
and two far detuning microwaves are used to realize the coupling between $|2\rangle$ and $|3\rangle$ through microwave Raman transitions.
(c) Experimental sequences for adiabatic passage without using the spin echo technology. In the beginning, an ion is cooled to the Doppler limit by an auto-modulated Doppler-cooling beam. The state is then initialized to $|1\rangle$, which is the highest energy eigenstate of $H(\bm{k})$
at a starting momentum for the adiabatic passage. Next, the Hamiltonian $H(\bm{k})$ is slowly tuned from the starting momentum to a final momentum following an adiabatic passage by modulating microwaves.
In the end, tomographic detections are performed under eight different bases to obtain the full density matrix of the final quantum state. }\label{fig:Experiment_system}
\end{figure*}

The experiments are performed in a surface-electrode ion trap, which is placed in a spherical octagon vacuum chamber
providing an ultra-high-vacuum (UHV) environment.
The surface trap has advantages in scalability based on the micro-fabrication technologies and realizations of more complicated quantum operations through individual controls in many different regions.
The surface trap was designed and fabricated in our team at CQI, IIIS, Tsinghua University. This version (see Fig.~\ref{fig:Chip_figure}) was completed in Sep., 2018. A linear slot in the center of the trap chip allows us to shine
laser beams on the ion along the direction perpendicular to the trap surface [see Fig.~\ref{fig:Chip_figure}(b)]. We can also shine laser beams on
the ion along the direction parallel to the trap surface.
A pair of planar RF electrodes are attached on both sides of the slot along the axial direction to generate a Paul trap in the radial direction.
The RF signal is generated by a signal generator at $20.27\,$MHz and then amplified by a commercial power amplifier, whose amplitude is controlled by a voltage variable attenuator (VVA) and fixed by an amplitude locking feedback.
The maximum radial secular frequency can reach $3\,$MHz. The breakdown voltage for the RF power is about $500\,$V.
The depth of a trap is typically $0.2\sim0.3\,$eV. The potential confinement strength and trap frequencies are optimized by calculating electrode configurations, including the width of the RF electrodes, gaps between electrodes, sizes and patterns of electrodes and applied voltages.

The axial confinement is provided by 20 pairs of segmented DC electrodes and one pair of global inner DC electrodes through applying
the voltage within $20\,$V; this allows for full controls of principal axis rotations and transport of ion chains
by varying the DC voltages. We usually apply asymmetric DC voltages along the radial direction to tilt the principle axes
in order to realize a best Doppler cooling along each direction.
Meanwhile, unequal voltages are applied to the inner DC rails to
reduce the micro motion effects. To minimize the heating effects and maximize our control ability of ions, we use the configurations so that the height of ions is about $100\mu$m above the surface. Note that
the potential field above the surface trap is calculated by a BEM simulation program written by us (https://github.com/zhaowending/BEM.git).

\subsection*{2. Laser and microwave operations and detection}

A $370\,$nm laser beam modulated by AOMs and EOMs are used for most operations on the ions, including the cooling, pumping and detection.
The relative detuning of  the beams (compared to the setting of laser device) are:
+110 MHz with AOM modulation for the
protection beam, 14.738 GHz sidebands with EOM modulation and +260 MHz with AOM modulation for the cooling beam,
2.105 GHz sidebands with EOM modulation and +272 MHz with AOM modulation for the pumping beam and +272 MHz with AOM modulation for the detection beam.
These $370\,$nm laser beams all drive the transition of $S_{1/2} \rightarrow P_{1/2}$ by coupling different energy levels. A field programmable gate array (FPGA) is used to control each of these signals so that we can switch AOMs and EOMs of these beams in microseconds.
In addition, the frequency of lasers is actively stabilized by a stable cavity system with a digital
Proportion Integration Differentiation feedback lock (PID-lock).
The relative frequency detuning of each beam is modulated and calibrated to a most effective setting before experiments.

In the experiment, the ${}^{171}$Yb atoms are generated by an oven, excited to the ${}^{1}P_{1}$ level by a $399\,$nm laser beam and then ionized.
These ions are further cooled into a Wigner crystal using a cooling beam. We also use a $935\,$nm laser beam to repump the ions in the ${}^{2}D_{3/2}$ state back to the ground state. See Fig.~\ref{fig:Experiment_system}(a) for schematics of the experimental setup. The three-level quantum system is defined in the $S_{1/2}$ manifold of a trapped ${}^{171}$Yb$^{+}$ ion as shown in Fig.~\ref{fig:Experiment_system}(b).
Here, the state $|1\rangle = |F=0,m_F=0 \rangle $ is a dark state, which can be prepared by a pumping beam.
The transitions between two of the levels $|1\rangle = |F=0,m_F=0\rangle $, $|2\rangle = |F=1,m_F = -1\rangle $ and $|3\rangle = |F=1,m_F =0 \rangle  $
are driven by microwaves as shown as Fig.~\ref{fig:Experiment_system}(b). The transition frequencies are $\omega_{13} = 2\pi \times 12.6428335\,$GHz and $\omega_{12} = \omega_{13} - 2\pi \times 11.5697\,$MHz. In order to control the amplitudes, frequencies and phases of microwaves and produce experimental sequences, we use a high frequency signal generator (keysight N5173B) and an arbitrary waveform generator (AWG) with $1.25\,$GHz sampling rate (Spectrum DN2.663) to generate microwaves with different frequencies.
These two signals generated by the two devices are mixed by an IQ mixer,
where the difference frequency sidebands are suppressed and the sum frequency sidebands are enhanced;
the latter are employed to drive transitions between two different energy levels in the ion.
All of the signal generators are synchronized by a $10\,$MHz reference rubidium clock (SRS FS725). The magnetic field of $B = 8.2640\,$G is generated by a Samarium cobalt magnet, which can generate highly stable magnetic fields due to its
insensitivity to temperatures.

\begin{figure}[htbp]
  \centering
  \includegraphics[width=0.8\linewidth]{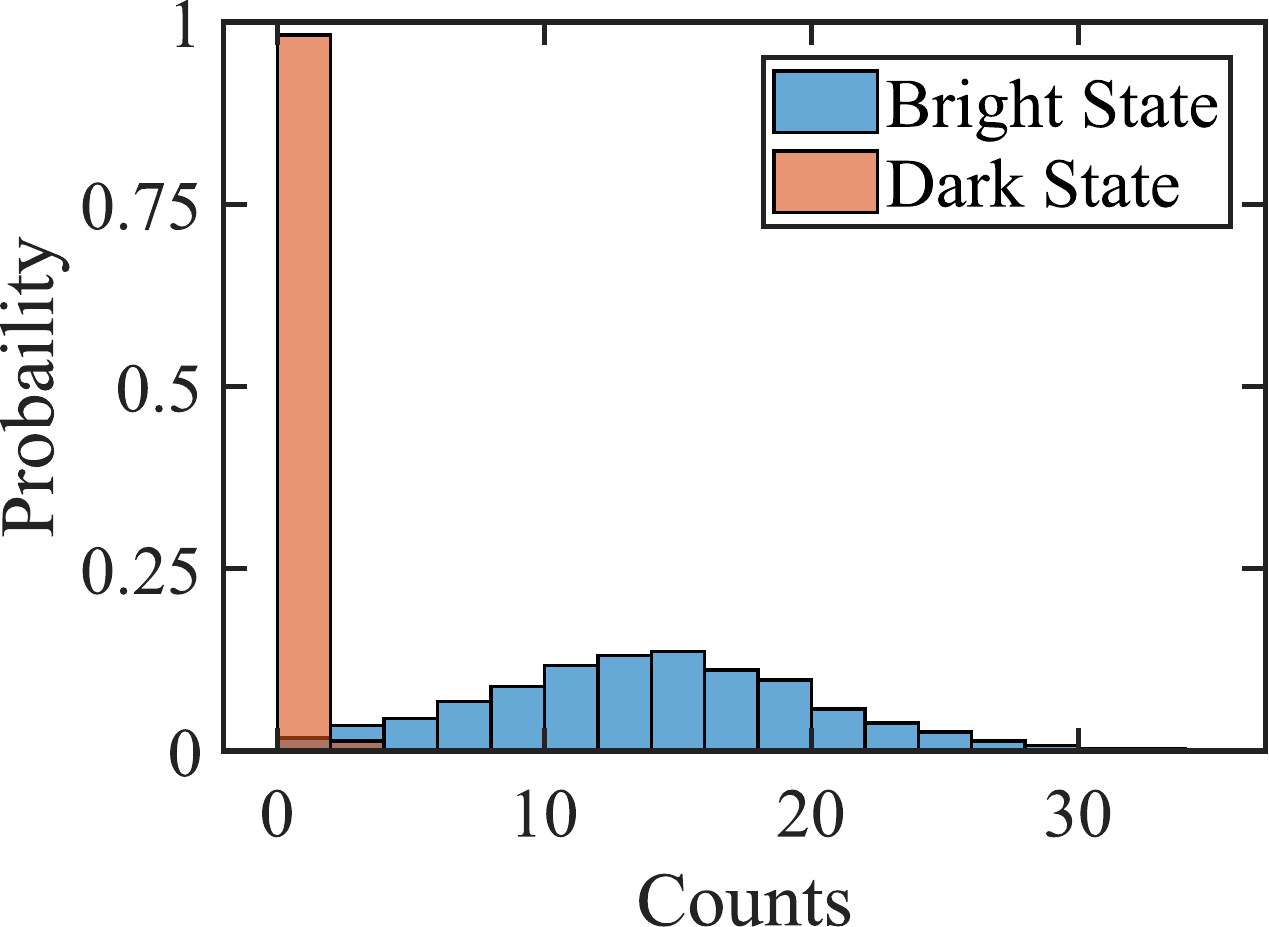}\\
  \caption{\textbf{Probability distributions of collected fluorescence photons for a $ 250\mu s$ detection of a single ion.}
  The red and blue histograms show the distributions obtained for a dark state ($|1\rangle$) and bright states (the average from $|2\rangle$ and $|3\rangle$), respectively. The two histograms follow the Poisson distributions with the expected value $N=0.06$ and $N = 13.31$, respectively. The detection fidelity is higher than $98\%$ by discriminating events with the collected photon counts larger than $1$ during a detection cycle that is typically $200\sim250 \mu s$. Here, the dark state on the level $|1\rangle$ is prepared by the optical pumping, and bright states on the levels $|2\rangle $ and $|3\rangle $ are prepared by microwave Rabi transitions. Since the infidelity of microwave Rabi operations is usually much smaller than the detection infidelity, we
  neglect its contributions when calculating the detection fidelity.}\label{fig:detect_fidelity}
\end{figure}

In measurements,
the ion's fluorescence from spontaneous emission is collected by a homemade objective with the numerical aperture $NA=0.33$ and imaged by a photo-multiplier tube (PMT) or an electron-multiplying charged couple device (EM-CCD). In this experiment, the fluorescence photon from a single ion is collected by a PMT, which is controlled by a FPGA.
The detection time and light intensity are optimized and calibrated each time before experiments to ensure a highest detection fidelity. The average probabilities for state detection errors are: $1.90\%$ for detecting a bright state as a dark state and  $0.56\% $ for detecting a dark state as a bright state,
namely, the average detection fidelity for the dark state and the bright state are $99.44\%$ and $98.10\%$, respectively.
The histogram of collected fluorescence counts is shown in Fig.~\ref{fig:detect_fidelity}.

\section*{\label{sec:SP_micro_operation}Appendix B: Microwave operations in a trapped ion system}

\setcounter{equation}{0}
\renewcommand{\theequation}{B\arabic{equation}}
\renewcommand{\theHequation}{B\arabic{equation}}
\renewcommand{\bibnumfmt}[1]{[#1]} \renewcommand{\citenumfont}[1]{#1}

In our trapped ion system, we use microwaves to drive transitions between two of the levels $|1\rangle$, $|2\rangle$ and $|3\rangle$.
The transition between $|1\rangle$ and $|2\rangle $ or between $|1\rangle$ and $|3\rangle $ is realized by near resonant microwaves
and the transition between $|2\rangle$ and $|3\rangle $ is realized by a pair of far detuning microwaves through microwave Raman transitions~\cite{2007Effective}.
Therefore, in the experiment, four microwave pulses with distinct frequencies are combined and applied to the ion as shown in
Fig.~\ref{fig:Experiment_system}(b).

%\subsection{Effective Hamiltonian of the qutrit system}
A qutrit system interacted with microwaves is described by the Hamiltonian
\begin{equation}
\hat{H} = \hat{H}_0 + \hat{H}_{AL},
\end{equation}
where the atomic part is
\begin{gather*}
 \hat{H}_0= \hbar (\omega_{hf} - \omega_z )|2\rangle \langle 2 | + \hbar(\omega_{hf} +\omega_q)|3\rangle \langle 3|
\end{gather*}
with $\omega_{hf}$ being the central transition frequency between $|1\rangle $ and $|3\rangle$, and $\omega_z$ ($\omega_q$) being
the frequency of the first-order (second-order) Zeeman energy determined by magnetic fields.
The interaction part due to the microwaves is
\begin{equation}\label{eq:H_AL_initial}
  \hat{H}_{AL}(t)= \sum_{n=1}^{4} \sum_{j=2,3} \hbar \Omega_{1j}^{(n)} \cos(\omega_n t + \phi_n)\sigma_{x}^{j}
\end{equation}
where $\hat{\sigma}_{x}^{(j)}=|1\rangle\langle j|+H.c.$, $\omega_n$ and $\phi_n$ are the frequency and initial phase of each microwave, and
$\Omega_{ij}^{(n)}$ is the Rabi frequency for the $|i\rangle \leftrightarrow |j\rangle $ transition driven by the $n$th
microwave, which depends on the microwave's polarization and strength. The four frequencies are given by
\begin{eqnarray}
  \omega_1 &=& \omega_{hf}-\omega_z+\delta_1  \\
  \omega_2 &=& \omega_{hf}+\omega_q+\delta_2  \\
  \omega_3 &=& \omega_{hf}-\omega_z-\Delta_1 \\
  \omega_4 &=& \omega_{hf}+\omega_q-\Delta_2,
\end{eqnarray}
where $\Delta_i $ is the detuning for the stimulated Raman transition and $\delta_i$  is the frequency shift used to compensate the AC stark shift.

We now write the Hamiltonian in the interaction picture,
$\hat{H}_{I} = (U_{1}^{\dag} \hat{H} U_1 - \hat{H}_0) $ where $U_1 = e^{-i\frac{\hat{H}0_0}{\hbar} t}$.
Using rotation wave approximations (RWA),
the interaction Hamiltonian can be re-written in the following form:

\begin{equation}\label{eq:H_i_final_form}
  \hat{H}_I(t) = \Sigma_{n=1}^{N} \hat{h}_n \exp(-i\omega_n t) + \hat{h}_n^{\dag} \exp(i\omega_n t),
\end{equation}
where $N$ is the total number of the harmonic terms making up the interaction Hamiltonian and $\omega_1 \leq \omega_2 \leq...\leq \omega_N$.
Following Ref.~\cite{2007Effective}, we obtain an effective Hamiltonian
\begin{equation}\label{eq:H_eff_cal}
  \hat{H}_{\textrm{eff}} = \Sigma_{m,n=1}^{N} \frac{1}{\hbar\overline{\omega}_{mn}} [\hat{h}_m^{\dag}, \hat{h}_n] e^{i(\omega_m- \omega_n)t},
\end{equation}
where $\overline{\omega}_{mn} $ is the harmonic average of $\omega_m$ and $\omega_n$, namely,
\begin{equation}\label{eq:omega_mn_average}
  \frac{1}{\overline{\omega}_{mn}} = \frac{1}{2} (\frac{1}{\omega_m} + \frac{1}{\omega_n} ).
\end{equation}

\begin{widetext}
In our system, we derive the effective Hamiltonian $\hat{H}_{\textrm{eff}} = \hat{H}_{st} + \hat{H}_{cp}$ in the interaction picture as
\begin{eqnarray}
 \hat{H}_{\textrm{eff}}=  {\left [ \begin{array}{ccc}
  H_{st,11} & \frac{\hbar\Omega_{12}^{(1)}}{2}e^{i\delta_1 t+i\phi_1} & \frac{\hbar\Omega_{13}^{(2)}}{2} e^{i\delta_2 t+i\phi_2} \\  \frac{\hbar\Omega_{12}^{(1)}}{2}e^{-i\delta_1 t-i\phi_1} & H_{st,22} & \frac{\hbar\Omega_{12}^{(3)}\Omega_{13}^{(4)}(\Delta_1+\Delta_2)}{8\Delta_1\Delta_2}e^{i(\phi_4-\phi_3)}e^{i(\Delta_1-\Delta_2)t} \\ \frac{\hbar\Omega_{13}^{(2)}}{2} e^{-i\delta_2 t-i\phi_2}  & \frac{\hbar\Omega_{12}^{(3)}\Omega_{13}^{(4)}(\Delta_1+\Delta_2)}{8\Delta_1\Delta_2}e^{i(\phi_3-\phi_4)}e^{i(\Delta_2-\Delta_1)t}  & H_{st,33}  \\
  \end{array}
   \right ]},
\end{eqnarray}
where the diagonal terms  $\hat{H}_{st}$ are
\begin{align}\label{eq:st11}
  H_{st,11}= &\frac{\hbar(\Omega_{12}^{(2)})^2 }{4(\omega_q+\omega_z+\delta_2)} - \frac{\hbar(\Omega_{12}^{(3)})^2}{4\Delta_1} -\frac{\hbar(\Omega_{12}^{(4)})^2}{4(\Delta_2-\omega_z-\omega_q)} \nonumber \\
  &- \frac{\hbar(\Omega_{13}^{(1)})^2}{4(\omega_z+\omega_q-\delta_1)} - \frac{\hbar(\Omega_{13}^{(3)})^2}{4(\omega_z+\omega_q+\Delta_1)}
  - \frac{\hbar(\Omega_{13}^{(4)})^2}{4\Delta_2}
\end{align}

\begin{equation}\label{eq:st22}
  H_{st,22}= -\frac{\hbar(\Omega_{12}^{(2)})^2 }{4(\omega_q+\omega_z+\delta_2)}+\frac{\hbar(\Omega_{12}^{(3)})^2}{4\Delta_1}+ \frac{\hbar(\Omega_{12}^{(4)})^2}{4(\Delta_2-\omega_z-\omega_q)}
\end{equation}

\begin{equation}\label{eq:st22}
  H_{st,33}=\frac{\hbar(\Omega_{13}^{(1)})^2}{4(\omega_z+\omega_q-\delta_1)}+ \frac{\hbar(\Omega_{13}^{(3)})^2}{4(\omega_z+\omega_q+\Delta_1)} + \frac{\hbar(\Omega_{13}^{(4)})^2}{4\Delta_2}.
\end{equation}

This Hamiltonian can be further transformed into a time-independent form in the rotating frame by $U$ through
\begin{equation}\label{eq:2nd_rotate_frame}
  \hat{H}_{\textrm{eff}}^{'} = U^{\dag} H_{\textrm{eff}} U - i\hbar U^{\dag} \frac{\partial U}{\partial t}
\end{equation}
so that the final time-independent Hamiltonian is
	\begin{eqnarray}\label{eq:Heff_end_Micro}
	   {H}_{\textrm{exp}}=  {\left [ \begin{array}{ccc}
   H_{st,11} & \frac{\hbar\Omega_{12}^{(1)}}{2}e^{i\phi_1} & \frac{\hbar\Omega_{13}^{(2)}}{2} e^{i\phi_2} \\  \frac{\hbar\Omega_{12}^{(1)}}{2}e^{-i\phi_1} & H_{st,22}-\hbar\delta_1 & \frac{\hbar\Omega_{12}^{(3)}\Omega_{13}^{(4)}(\Delta_1+\Delta_2)}{8\Delta_1\Delta_2}e^{i(\phi_4-\phi_3)} \\ \frac{\hbar\Omega_{13}^{(2)}}{2} e^{-i\phi_2}  & \frac{\hbar\Omega_{12}^{(3)}\Omega_{13}^{(4)}(\Delta_1+\Delta_2)}{8\Delta_1\Delta_2}e^{i(\phi_3-\phi_4)} & H_{st,33}-\hbar\delta_2
   \end{array}
   \right ]}.
	\end{eqnarray}
\end{widetext}

The Euler Hamiltonian can thus be experimentally engineered by tuning the initial phase $\phi_i$, the
detuning and microwave amplitudes. Note that the values of $\Delta_i$ ($i=1,2$) are chosen to be servals times of the values of the Rabi frequencies
to realize the Raman transitions.
The Rabi frequencies of each transition with different power are all calibrated before experiments. In order to realize a faster and more effective adiabatic evolution, the strength of each transition is balanced via changing the directions of polarization of the microwave horn and magnetic field.

In addition, the maximum value of the coupling between $|2\rangle $ and $|3\rangle $ that we can reach in the experiment is much smaller than the values of the couplings between $|1\rangle $ and $|2\rangle $ or $|1\rangle $ and $|3\rangle $, since the former coupling is realized through microwave Raman transitions. When $H_{23}$ is much larger than $H_{12}$ or $H_{13}$ in $H(\bm{k})$, the energy scale $c$ is very small so that a much longer period of
time is required for the adiabatic evolution. To overcome the difficulty, we apply a $\pi$-rotation between $|1\rangle$ and $|2\rangle$ (or $|1\rangle$ and $|3\rangle$) at an appropriate moment during the adiabatic passage to change the basis for the qutrit system. The resultant effective Hamiltonian relative to the new basis has a smaller entry $H_{23}$, so that a larger value of the coefficient $c$ is obtained, which effectively reduces the time for the adiabatic evolution.
Meanwhile, we continue the adiabatic passage for the Hamiltonian relative to the new basis and apply another $\pi$-rotation before detections for quantum state tomography.

In the experiment, we implement the Hamiltonian
\begin{equation}
{H}_{\textrm{exp}} (\bm{k}) = c [ H(\bm{k}) - b {I_3} ],
\end{equation}
which has the same eigenstates as $H(\bm{k})$ and thus is topologically equivalent to $H(\bm{k})$. Here, $I_3$ is the $3\times 3$ identity matrix. $c$ is a numerically computed maximum value for our accessible system parameters in each $\bm k$ calculated via the fmincon function in MATLAB so that the energy gap takes a maximum value.

\section*{\label{sec:tomography_detect}Appendix C: Quantum state tomography}

\setcounter{equation}{0}
\renewcommand{\theequation}{C\arabic{equation}}
\renewcommand{\theHequation}{C\arabic{equation}}
\renewcommand{\bibnumfmt}[1]{[#1]} \renewcommand{\citenumfont}[1]{#1}

\begin{figure*}
  \centering
  \includegraphics[width=0.8\textwidth]{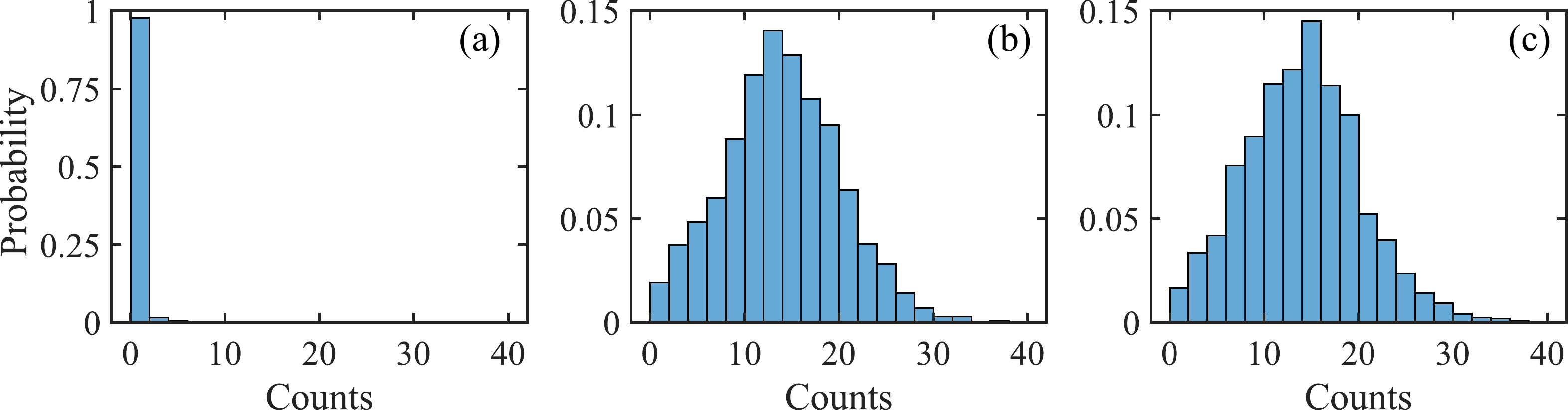}\\
  \caption{\textbf{PMT photon counts in a $250\mu s $ detection.} (a), (b) and (c) PMT photon count histograms for
   the levels $|1\rangle$, $|2\rangle$ and $|3\rangle$, respectively.
   The corresponding average counts are $N_1=0.06$, $N_2= 13.28$ and $N_3=13.34$, respectively. }\label{fig:Counts_each_level}
\end{figure*}

In this appendix, we will present how to measure a quantum state in the qutrit system through quantum state tomography~\cite{2002Qudit, 2019Machine}.
The density matrix can be written in terms of eight real parameters as
\begin{gather}\label{eq:define_density_matrix}
 \rho =\begin{pmatrix} a  & c+id & g+ih \\  c-id & b & e+if \\ g-ih & e-if & 1-a-b \\   \end{pmatrix}.
\end{gather}
The probability $a$ on the level $|1\rangle$ in the density matrix can be measured directly
using threshold counts to distinguish between the bright and dark states. However, we cannot distinguish whether the state is populated on $|2\rangle $ or  $|3\rangle$ based on the method. To obtain all the eight parameters, we perform measurements under eight different bases realized by applying $\pi$ or $\pi/2$ or both rotations on Bloch spheres before detection
as listed in Table~\ref{table:tomography_measure}.
The state probabilities in terms of the parameters are determined by the average fluorescence counts collected in detections over thousands of repeated experiments.
In state detections, we apply a $1\mu W$ detection beam for about $250\mu s$ to generate fluorescence, which is collected by a PMT.
In order to keep a high fidelity, the detection beam and the PMT are both calibrated and optimized based on the detection fidelity everyday before experiments.

\begin{table}[htbp]
\centering
\caption{\textbf{Microwave operations of eight tomographic detections.} $\pi_{\sigma}$($\pi_{\sigma}/2$) ($\sigma=X,Y$)
 refer to a $\pi$ ($\pi/2$) rotation on the Bloch sphere about the $\sigma$ axis, which is implemented by microwave Rabi operations.
 The value $P_1$ is the probability on the dark state ( $|1\rangle $ state) after the corresponding microwave rotations.}
\label{constant}
\begin{tabular}{c|c|c|c}
\toprule
   $Index$  & $|1\rangle \leftrightarrow |2\rangle $ & $|1\rangle \leftrightarrow |3\rangle $ & $P_1$ \\
 \midrule
  $1.$ & $-$  &  $-$  &  $a$ \\
  $2.$ & $-\pi_{Y}$  &  $-$  &  $b$ \\
  $3.$ & $\pi_{X}/2$  &  $-$  &  $\frac{a+b}{2} -d $ \\
  $4.$ & $-\pi_{Y}/2$  &  $-$  &  $\frac{a+b}{2} + c$ \\
  $5.$ & $-$  &  $\pi_{X}/2 $  &  $\frac{1-b}{2} -h$ \\
  $6.$ & $-$  &  $-\pi_{Y}/2$  &  $\frac{1-b}{2} + g$  \\
  $7.$ & $\pi_{X}/2$  &  -$\pi_{Y}$  &  $\frac{1-a}{2} +f $ \\
  $8.$ & $-\pi_{Y}/2$  &  -$\pi_{Y}$  &  $\frac{1-a}{2} +e$ \\
  \bottomrule
\end{tabular}
\label{table:tomography_measure}
\end{table}

To ensure that the measured density matrix is Hermitian and positive semi-positive, we utilize
the maximum likelihood estimation~\cite{2001Measurement,2019Machine}.
The physical density matrix can be defined as
\begin{equation}\label{eq:rho_T_define}
  \hat{\rho}_{T}(t) = \hat{T}^{\dag}(t) \hat{T}(t)/ \mathrm{Tr}[\hat{T}^{\dag}(t)\hat{T}(t)],
\end{equation}
where $\hat{T}(t)$ is a tridiagonal matrix depending on a set of real parameters $t=\{t_1, t_2, \cdots, t_9\}$:
\begin{gather}\label{eq:define_Matrix_T}
 \hat{T}(t) =\begin{pmatrix} t_1  & 0 & 0 \\  t_4 + it_5 & t_2 & 0 \\ t_8+it_9 & t_6+it_7 & t_3 \\   \end{pmatrix}.
\end{gather}

Assuming that the noise of measurement results obeys a Gaussian distribution,
the probability to obtain a set of measurement data $\{n_1,n_2,\cdots,n_8\}$ is
\begin{equation}\label{eq:Probability_8counts_set}
  P(n_1, n_2, \cdots, n_8) = \prod_{i=1}^{8} \exp \left[-\frac{(n_i-\bar{n}_i)^2}{2\sigma_i^2}\right],
\end{equation}
where $n_i$ is the measured value for the photon count with the expected value $\bar{n}_i$
and $\sigma_i$ is the standard deviation for the $i$th measurement.
Given a density matrix $\hat{\rho}_{T}(t)$ in terms of a set of real parameters $t=\{t_1, t_2, \cdots, t_9\}$, one can calculate $\bar{n}_i$ via
\begin{equation}\label{eq:define_bar_ni}
  \bar{n}_i (t_1,t_2,\cdot \cdot \cdot ,t_9) = N_1 \bar{P}_{1}^i(t) + N_2\bar{P}_{2}^i(t) + N_3\bar{P}_{3}^i(t),
\end{equation}
where $\bar{P}_{j}^i(t)$ ($j=1,2,3$) denotes the probability on the state $|j\rangle$ for the $i$th ($i=1,2,\cdots,8$) measurement
computed based on
the density matrix $\hat{\rho}_{T}(t)$, and $N_j$ denotes the average fluorescence count if the measured state is $|j\rangle$.
The typical PMT counts in a $250\mu s$ detection for each energy level is $N_1 = 0.06$, $N_2 = 13.28 $ and $ N_3 = 13.34$
(also see Fig.~\ref{fig:Counts_each_level} for the distribution of PMT counts for each energy level).

With the experimentally measured PMT counts $n_j$ ($j=1,2,\cdots,8$), one can determine an optimum set of $\{t_1,t_2,\cdots,t_9\}$
in the density matrix $\hat{\rho}_{T}(t)$ to maximize the $P$. This is equivalent to finding the minimum of the function $ l(t) $:
\begin{equation}\label{eq:negative_logarithm_l}
  l(t_1,t_2, \cdots, t_9) = \sum_{i=1}^{8} \frac{[\bar{n}_i(t_1,t_2, \cdots, t_9)- n_i]^2}{2\bar{n}_i(t_1,t_2, \cdots, t_9)}.
\end{equation}
Using numerical optimization tools (via the fmincon function in MATLAB), we determine the optimum set of $\{t_1,t_2,\cdots,t_9\}$ to minimize $l(t_1,t_2, \cdots, t_9)$
and obtain the best estimate for the physical density matrix $\rho_T(t_1,t_2,\cdots,t_9)$.

\section*{Appendix D: Analysis of fidelity and errors}

\setcounter{equation}{0}
\renewcommand{\theequation}{D\arabic{equation}}
\renewcommand{\theHequation}{D\arabic{equation}}
\renewcommand{\bibnumfmt}[1]{[#1]} \renewcommand{\citenumfont}[1]{#1}

\begin{figure}
  \centering
  \includegraphics[width=\linewidth]{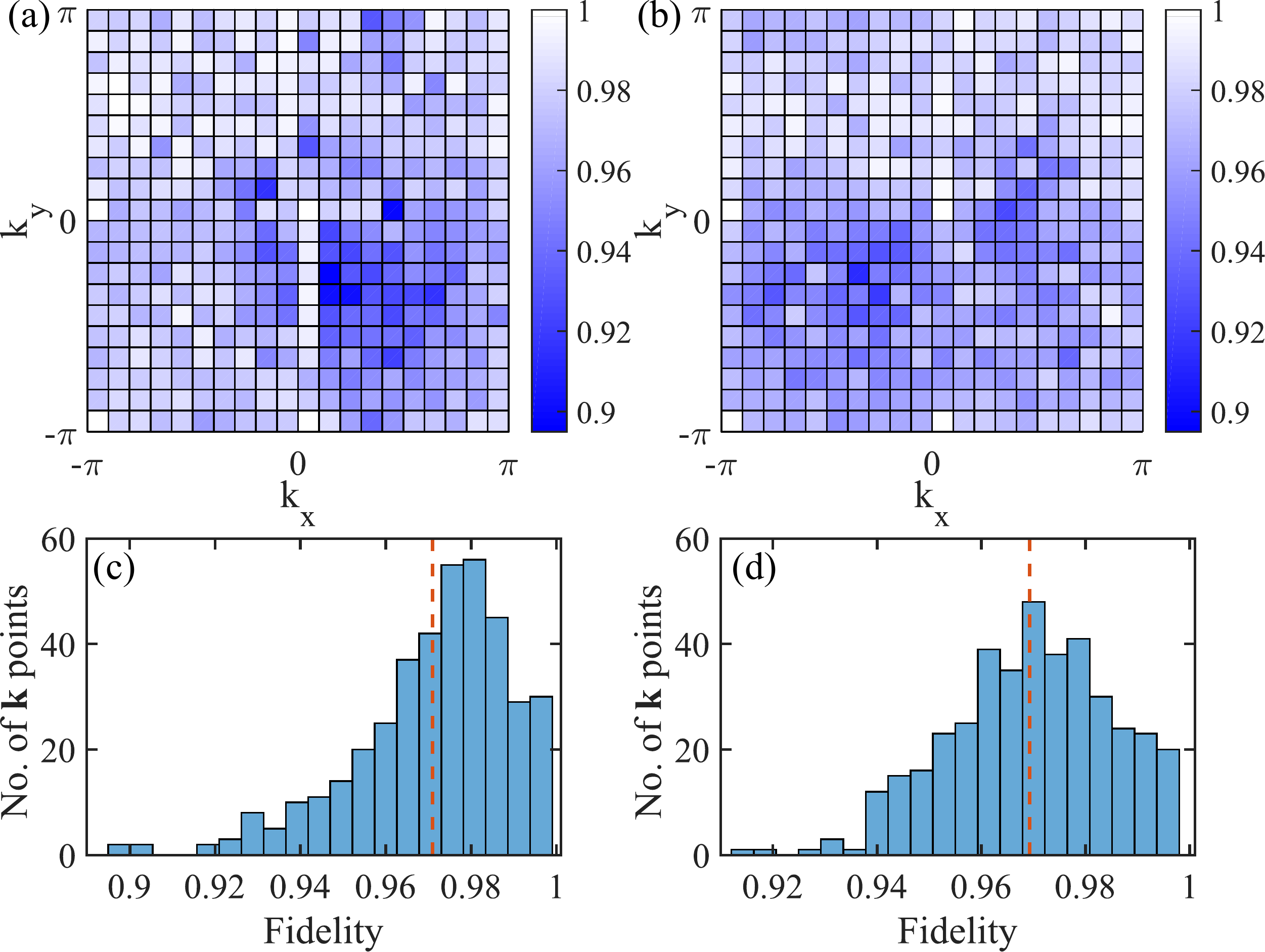}\\
  \caption{ \textbf{Fidelity for experiment results}. (a-b) Fidelity for the measured states with respect to $k_x$ and $k_y$ ($21\times 21$
  momentum $\bm k$ points).
  In (a), $m=1$ and in (b), $m=3$, corresponding to topologically nontrivial and trivial phases, respectively.
  Their corresponding distributions are further plotted as a histogram in (c) and (d), respectively.
  The average fidelities are $F=0.971$ and $F =0.9693$ for the nontrivial and trivial phases, respectively, as highlighted by the red dashed line.}\label{fig:Fidelity_4plot_end}
\end{figure}

The fidelity at each ${\bm k}$ is calculated by
\begin{equation}\label{eq:define_fidelity}
  F(\bm{k}) = \langle \psi(\bm{k}) | \rho(\bm{k}) | \psi(\bm{k}) \rangle,
\end{equation}
where $\psi(\bm{k})$ is the theoretically obtained state, and $\rho(\bm{k})$ is the measured density matrix optimized by
the maximum likelihood estimation. The optimization by maximum likelihood estimation can improve the fidelity by $2\% \sim 4\%$. In Fig.~\ref{fig:Fidelity_4plot_end}, we plot the fidelity for the measured states at each momentum; these states are also used to evaluate the Euler class, Wilson loop spectra and entanglement spectra shown in Fig. 2 and Fig. 3
in the main text. The average fidelities are $F=97.38\%$ and $F =97.07\%$ for the topologically nontrivial and trivial phases, respectively.

To identify the band topology of the Euler Hamiltonian, we need to transform the measured density matrix $\rho$ into a real state $|\psi\rangle=(\alpha, \beta, \gamma)^T$ by maximizing the function
 \begin{equation}\label{eq:near_state}
   f (\alpha, \beta, \gamma)= |\langle \psi | \rho | \psi \rangle |
\end{equation}
so that $| \psi \rangle$ is the closest real state to the measured density matrix $\rho$.
We find that the average fidelities between the real state $|\psi\rangle$ and the density matrix $\rho$ are $97.38\%$ and $97.07\%$ when $m=1$ (nontrivial) and $m=3$ (trivial), respectively. We also find that the fidelities between the closest complex pure states and the density matrix are $97.52\%$ and $97.27\%$, respectively. We see that the infidelity resulted from the restriction to a real state rather than a general complex pure state
in the above maximization is below $1\%$, suggesting that the measured density matrix may not correspond to a pure state due to decoherence and detection errors.

The infidelity may arise from the detection infidelity of the dark state or the bright state, microwave pulse errors caused by nonlinear effects of experimental equipment and environment fluctuations. In addition, decoherence is also a factor for infidelity when the microwave operations take a long period of time. The coherence time is typically $600\mu s$ in our trapped-ion system, which is mainly restricted by the Zeeman state. In the experiment, we perform calibrations and optimize the experiment setups per hour in order to obtain a high fidelity.

\section*{Appendix E: The Euler class for a three-band Euler insulator}

\setcounter{equation}{0}
\renewcommand{\theequation}{E\arabic{equation}}
\renewcommand{\theHequation}{E\arabic{equation}}
\renewcommand{\bibnumfmt}[1]{[#1]} \renewcommand{\citenumfont}[1]{#1}

In this appendix, we will present a proof (which is equivalent to the proof in Ref.~\cite{Slager2020NP})
of the statement that the general formula for the Euler class can be reduced to a formula for a winding number (Pontryagin number) in a
three-band Euler insulator.
We further demonstrate that the Euler class as a winding number can be evaluated as a Chern number of a derived Chern insulator.

We first show the
equivalence between the integrands of the two forms of the Euler class.
Here, we consider a three-band Euler Hamiltonian in momentum space
\begin{equation}
H(\bm{k})=2\bm{n}(\bm{k})\cdot\bm{n}(\bm{k})^{T}-I_3,
\end{equation}
which has three eigenstates represented by real unit vectors as $|u_i(\bm{k})\rangle = \bm{u}_{i}(\bm{k})$ with $i=1,2,3$.
The highest energy eigenstate can be written in terms of two other eigenstates as $\bm{u}_{3}(\bm{k}) = \bm{n}(\bm{k})= \bm{u}_{1}(\bm{k}) \times \bm{u}_{2}(\bm{k})$. We now derive that the integrand for the Euler class in Eq. (4) in the main text is equal to the integrand in Eq. (3) as follows:
\begin{widetext}
\begin{align}
\bm{n}\cdot(\partial_{k_x} \bm{n} \times \partial_{k_y} \bm{n})
&=\bm{u}_3 \cdot (\partial_{k_x} \bm{u}_3 \times \partial_{k_y} \bm{u}_3) \nonumber \\
&=\bm{u}_3 \cdot (\partial_{k_x} (\bm{u}_1 \times \bm{u}_2) \times \partial_{k_y} \bm{u}_3 ) \nonumber \\
&=(\bm{u}_3 \times \partial_{k_x} (\bm{u}_1 \times \bm{u}_2)) \cdot \partial_{k_y} \bm{u}_3  \nonumber \\
&= (\bm{u}_3 \times (\partial_{k_x}\bm{u}_1 \times \bm{u}_2 + \bm{u}_1 \times \partial_{k_x}\bm{u}_2)) \cdot \partial_{k_y} \bm{u}_3 \nonumber \\
&=(-(\bm{u}_3\cdot \partial_{k_x}\bm{u}_1)\bm{u}_2 + (\bm{u}_3\cdot \partial_{k_x}\bm{u}_2)\bm{u}_1)
\cdot \partial_{k_y} \bm{u}_3 \nonumber \\
&=-(\bm{u}_3\cdot \partial_{k_x}\bm{u}_1)(\bm{u}_2\cdot \partial_{k_y}\bm{u}_3)
+ (\bm{u}_3\cdot \partial_{k_x}\bm{u}_2)(\bm{u}_1\cdot \partial_{k_y}\bm{u}_3) \nonumber \\
&=(\bm{u}_3\cdot \partial_{k_x}\bm{u}_1)(\bm{u}_3\cdot \partial_{k_y}\bm{u}_2)
- (\bm{u}_3\cdot \partial_{k_x}\bm{u}_2)(\bm{u}_3\cdot \partial_{k_y}\bm{u}_1) \nonumber \\
&=\sum_{i=1}^{3}(\bm{u}_i\cdot \partial_{k_x}\bm{u}_1)(\bm{u}_i\cdot \partial_{k_y}\bm{u}_2)
- \sum_{i=1}^{3}(\bm{u}_i\cdot \partial_{k_x}\bm{u}_2)(\bm{u}_i\cdot \partial_{k_y}\bm{u}_1) \nonumber \\
&=\sum_{i=1}^{3} \langle \partial_{k_x} u_1|u_i\rangle\langle u_i |\partial_{k_y} u_2\rangle
- \sum_{i=1}^{3} \langle \partial_{k_y} u_1|u_i\rangle\langle u_i |\partial_{k_x} u_2\rangle \nonumber \\
&=\langle \partial_{k_x} u_1|\partial_{k_y} u_2\rangle
- \langle \partial_{k_y} u_1|\partial_{k_x} u_2\rangle.
\end{align}
\end{widetext}
In the derivation, we have used the orthogonality and completeness properties for $\bm{u}_{1,2,3}$
and the properties that $\bm{u}_i \cdot \partial_{k_{\mu}} \bm{u}_i = 0$ and
$\bm{u}_i \cdot \partial_{k_{\mu}} \bm{u}_j = -\bm{u}_j \cdot \partial_{k_{\mu}} \bm{u}_i$ for $i\neq j$.
Therefore, in a three-band Euler insulator, the Euler class can also be evaluated by the
winding number of $\bm{n}(\bm{k})$.

With the measured $\bm{n}(\bm{k})$ in the discretized Brillouin zone, we can numerically calculate the winding number
based on Eq. (4). We find that $\xi=1.90$ in the topological phase; the deviation from the quantized value of $2$
arises from numerical errors. In fact, we can map the measured $\bm{n}(\bm{k})$ over the 2D Brillouin zone to a two-band Hamiltonian
$H_{C}(\bm{k})= \bm{n}(\bm{k}) \cdot \bm{\sigma}$ (with two eigenstates $|\psi_{1,2}(\bm{k})\rangle$ corresponding to
eigenvalues $E_{1,2}=\mp 1$). The Chern number for the occupied band can be computed efficiently even for a
discretized Brillouin zone with finite grid points~\cite{2005Chern}. We find that the Chern number based on the
experimentally measured $\bm{n}(\bm{k})$ is $1$ so that $\xi=2$.

We also provide a proof for the equivalence of the Chern number $C=\frac{1}{2\pi} \int \mathrm{d}^2 \bm{k} \mathcal{F}_{xy}$
and the winding number of $\bm{n}(\bm{k})$ as follows:
\begin{align}
\mathcal{F}_{xy}&= i(\langle \partial_{k_x} \psi_1 |\partial_{k_y}\psi_1\rangle
- \langle \partial_{k_y}\psi_1| \partial_{k_x} \psi_1\rangle) \nonumber \\
&= -2\mathrm{Im} \sum_{i=1}^{2} \langle \partial_{k_x} \psi_1 |\psi_i\rangle\langle\psi_i |\partial_{k_y}\psi_1\rangle \nonumber \\
&=-2\mathrm{Im} \langle \partial_{k_x} \psi_1 |\psi_2\rangle\langle\psi_2 |\partial_{k_y}\psi_1\rangle \nonumber \\
&=-2\mathrm{Im} \frac{\langle \psi_1 |\partial_{k_x}H | \psi_2 \rangle \langle \psi_2 |\partial_{k_y}H | \psi_1 \rangle}{(E_1-E_2)^2} \nonumber \\
&=-\frac{1}{2} \mathrm{Im} \sum_{i=1}^{2} \langle \psi_1 |\partial_{k_x}H | \psi_i \rangle \langle \psi_i |\partial_{k_y}H | \psi_1 \rangle \nonumber \\
&=-\frac{1}{2} \mathrm{Im} \langle \psi_1 |(\partial_{k_x}H) (\partial_{k_y}H) | \psi_1 \rangle \nonumber \\
&=-\frac{1}{2} \mathrm{Im} \langle \psi_1 |(\partial_{k_x}\bm{n}\cdot \bm{\sigma}) (\partial_{k_y}\bm{n}\cdot \bm{\sigma}) | \psi_1 \rangle \nonumber \\
&=-\frac{1}{2} \mathrm{Im} \langle \psi_1 |(\partial_{k_x}\bm{n}\cdot \partial_{k_y}\bm{n}
+ i(\partial_{k_x}\bm{n}\times \partial_{k_y}\bm{n})\cdot\bm{\sigma}) | \psi_1 \rangle \nonumber \\
&=-\frac{1}{2}\langle \psi_1 | \bm{\sigma} | \psi_1 \rangle \cdot (\partial_{k_x}\bm{n}\times \partial_{k_y}\bm{n}) \nonumber \\
&=\frac{1}{2} \bm{n} \cdot (\partial_{k_x}\bm{n}\times \partial_{k_y}\bm{n}).
\end{align}
In the derivation, we have used the properties that $\mathrm{Im} \langle \partial_{k_x} \psi_1 |\psi_1\rangle\langle\psi_1 |\partial_{k_y}\psi_1\rangle =0$,
$\mathrm{Im} \langle \psi_1 |\partial_{k_x}H | \psi_1 \rangle \langle \psi_1 |\partial_{k_y}H | \psi_1 \rangle =0$,
$\langle \partial_{k_x} \psi_1|\psi_2\rangle = \langle \psi_1 |\partial_{k_x}H | \psi_2 \rangle/(E_1-E_2)$.
It can be seen that the winding number $\frac{1}{4\pi}\int \mathrm{d}^2 \bm{k} [\bm{n} \cdot (\partial_{k_x}\bm{n}\times \partial_{k_y}\bm{n})]$
is equal to the Chern number.

\section*{Appendix F: The fragile band topology for topological Euler insulators}

\setcounter{equation}{0}
\renewcommand{\theequation}{F\arabic{equation}}
\renewcommand{\theHequation}{F\arabic{equation}}
\renewcommand{\bibnumfmt}[1]{[#1]} \renewcommand{\citenumfont}[1]{#1}

In this appendix, we will illustrate that a three-band topological Euler insulator possesses a fragile band topology, namely,
it becomes topologically trivial when additional trivial bands are coupled to the occupied bands~\cite{Po2018PRL}.

We start from the flattened Euler Hamiltonian with three bands
\begin{equation}\label{Euler3flat}
H_3(\bm{k})=2\bm{n}_3(\bm{k})\cdot\bm{n}_3(\bm{k})^{T}-I_3,
\end{equation}
where $\bm{n}_3(\bm{k})$ is a real unit vector at each momentum $\bm{k}=(k_x,k_y)$ as
\begin{align}\label{Euler3flatnk}
\bm{n}_3(\bm{k})
&=\left(n_x(\bm{k}),n_y(\bm{k}),n_z(\bm{k})\right)^T \nonumber \\
&=\frac{1}{\mathcal{N}}\left( m-\cos(k_x)-\cos(k_y),\sin(k_x),\sin(k_y) \right)^T.
\end{align}
This Euler insulator model is topologically nontrivial with Euler class $\xi=2$ for $0<|m|<2$.
In mathematics, the band topology of a real three-band Euler Hamiltonian with $C_2 \mathcal{T}$ symmetry is classified by the second homotopy group
$\pi_2(O(3)/[O(2)\times O(1)])=\pi_2(\mathbb{R}P^2)=\mathbb{Z}$~\cite{Slager2020PRB}.

While the topological Euler insulator has nontrivial Euler class for two occupied bands, it becomes topologically trivial
upon adding an additional trivial band into the occupied space.
This can be explained by the second homotopy group $\pi_2(O(4)/[O(3)\times O(1)])=0$~\cite{Slager2020PRB}.
We will explicitly demonstrate this trivialization for the three-band Euler Hamiltonian in Eq.~(\ref{Euler3flat}).
We take $m=1$ in the topological phase of the Euler insulator, and add a trivial band with onsite energy $E_4=-1$ to the three-band Hamiltonian.
Then we get a four-band Hamiltonian as
\begin{equation}\label{Euler4flat}
H_4(\bm{k})=
\left(
  \begin{array}{cc}
    H_3(\bm{k}) & 0 \\
    0 & -1 \\
  \end{array}
\right),
\end{equation}
which has an unoccupied band at energy $E=1$ and three occupied bands at energy $E=-1$.
The Hamiltonian $H_4(\bm{k})$ can also be written as
\begin{equation}
H_4(\bm{k})=2\bm{n}_4(\bm{k})\cdot \bm{n}_4(\bm{k})^T - I_4,
\end{equation}
where $\bm{n}_4(\bm{k})=(\bm{n}_3(\bm{k});0)$ with $\bm{n}_3(\bm{k})$ in Eq.~(\ref{Euler3flatnk}).
Now we construct a continuous path of Hamiltonian $H_4(\bm{k},s)$ parameterized by $s\in [0,1]$
that connects the Hamiltonian $H_4(\bm{k},s=0)=H_4(\bm{k})$ to a topologically trivial Hamiltonian $H_4(\bm{k},s=1)$ as follows,
\begin{align}\label{Euler4flatpath}
&H_4(\bm{k},s)=2\bm{n}_4(\bm{k},s)\cdot \bm{n}_4(\bm{k},s)^T - I_4 \nonumber \\
=&\cos^2(\pi s/2) H_4(\bm{k}) +
\left(
  \begin{array}{cc}
    -\sin^2(\pi s/2)I_3 & \sin(\pi s) \bm{n}_3(\bm{k}) \\
    \sin(\pi s) \bm{n}_3(\bm{k})^T & \sin^2(\pi s/2) \\
  \end{array}
\right)
,
\end{align}
where $\bm{n}_4(\bm{k},s)=(\cos(\pi s/2)\bm{n}_3(\bm{k}); \sin(\pi s/2))$ with $s \in [0,1]$.
We can see that during this continuous deformation of the Hamiltonian, the gap between the unoccupied band and occupied bands remains open
and the relevant symmetry $C_2 \mathcal{T}$ is maintained.
Therefore, the four-band Hamiltonian $H_4(\bm{k})$ (\ref{Euler4flat}) is adiabatically connected to a topologically trivial Hamiltonian
$H_4(\bm{k},s=1)=\mathrm{diag}(-1,-1,-1,1)$ with localized Wannier functions.

\begin{figure*}[t]
\centering
\includegraphics[width=0.8\textwidth]{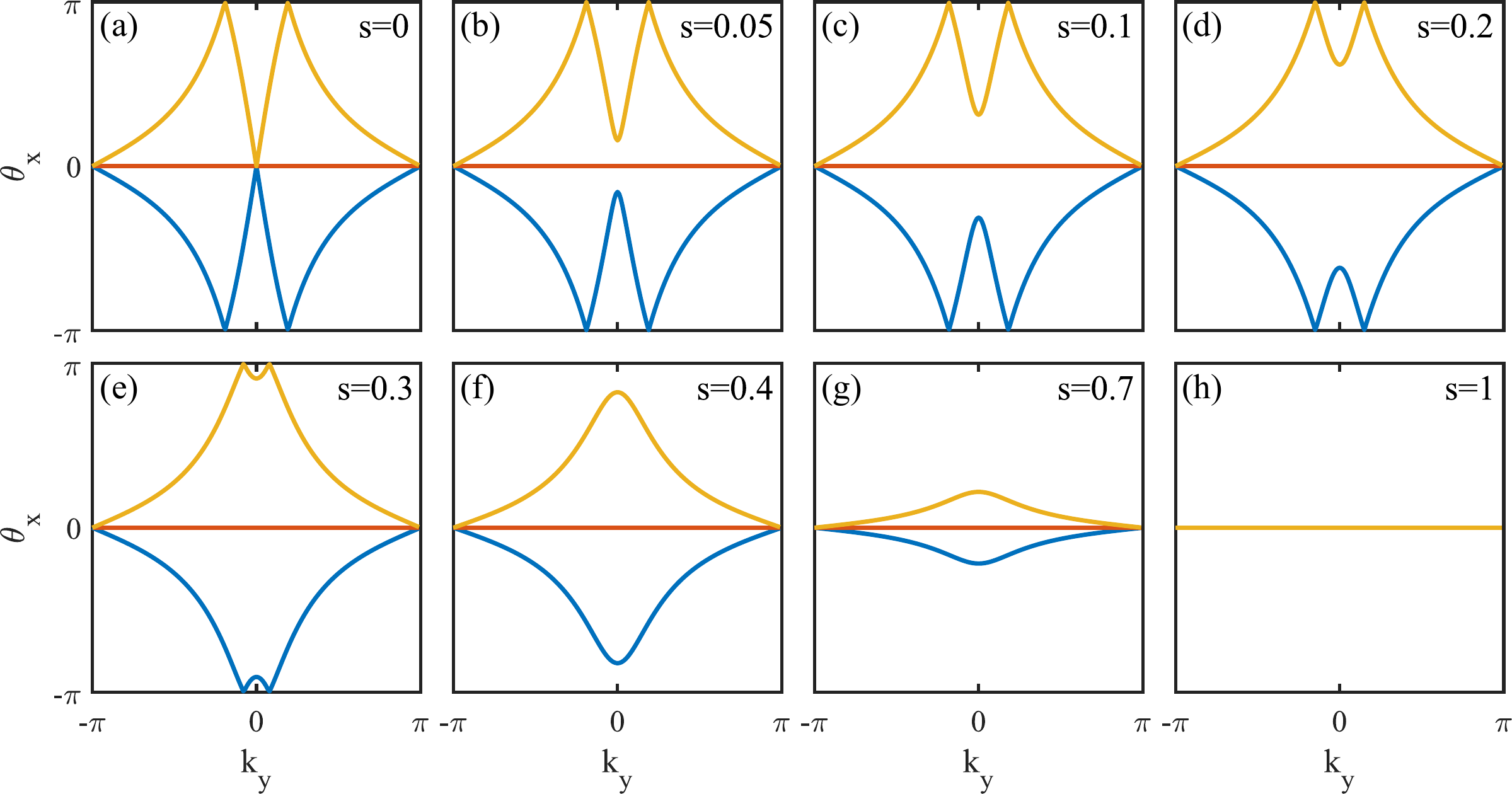}
\caption{\textbf{Fragile topology revealed by the Wilson loop spectrum.}
The evolution of the Wilson loop spectrum $\theta_x(k_y)$ over the three occupied bands for the Hamiltonian $H_4(\bm{k},s)$~(\ref{Euler4flatpath})
as the parameter $s$ changes from zero to one, revealing that
the nontrivial winding pattern of the Wilson loop spectrum can be removed by continuously deforming the Hamiltonian
without closing the energy gap between the occupied and unoccupied energy bands.
The Wilson loop spectrum $\theta_y(k_x)$ has similar behaviours.
}
\label{fig_FragileWilsonLoopX}
\end{figure*}

The fragile topology can also be characterized by the Wilson loop spectra.
For a three-band topological Euler insulator, the Wilson loop spectra over the two-band occupied space have nontrivial spectral flows
as shown in the main text, which indicate the obstruction to continuously deforming the real-space wave functions to localized Wannier functions.
When the two occupied bands are coupled to a third trivial band, the Wilson loop spectra will become gapped and have no nontrivial winding.
In Fig.~\ref{fig_FragileWilsonLoopX}, we plot the evolution process of the $x$-directed Wilson loop spectrum $\theta_x(k_y)$ for the four-band Hamiltonian
on the continuous path $H_4(\bm{k},s)$ (\ref{Euler4flatpath}) [the $y$-directed Wilson loop spectrum $\theta_y(k_x)$ is similar].
Figure~\ref{fig_FragileWilsonLoopX} clearly shows that as the parameter $s$ deviates from $0$, the Wilson loop becomes gapped at the triple point with $\theta_x=0$ at $k_y=0$,
and the gap becomes larger as $s$ increases.
Meanwhile, the two degenerate points with $\theta_x=\pi$ move closer and annihilate in pairs
so that the Wilson loop spectrum unwinds, and finally the three eigenvalues all approach the trivial value $\theta_x=0$
corresponding to the localized Wannier centers for an atomic insulator.

\section*{Appendix G: The entanglement spectra for topological Euler insulators}

\setcounter{equation}{0}
\renewcommand{\theequation}{G\arabic{equation}}
\renewcommand{\theHequation}{G\arabic{equation}}
\renewcommand{\bibnumfmt}[1]{[#1]} \renewcommand{\citenumfont}[1]{#1}

In this appendix, we will study the entanglement spectra (ES) for the three-band Euler insulator.
For a free-fermion system described by $\hat{H}= \sum_{ij} \hat{c}_i^{\dagger} H_{ij} \hat{c}_j$, we can compute the single-particle entanglement spectra,
which can exhibit more robust nontrivial features than those for physical boundaries in topological band insulators~\cite{Fidkowski2010PRL,Turner2010PRB,Hughes2011PRB}.
Let us consider a bipartition of the system into two subsystems denoted by $S$ and its complement $\overline{S}$.
The reduced density matrix for the subsystem $S$ of the many-body ground state can be written as
\begin{equation}
\hat{\rho}_S=\mathrm{Tr}_{\overline{S}} |\Psi_G \rangle\langle \Psi_G |=\frac{1}{Z} e^{-\hat{H}_E},
\end{equation}
where $|\Psi_G \rangle$ is the many-body ground state of the system,
and we define the Hermitian operator $\hat{H}_E$ as the entanglement Hamiltonian with $Z=\mathrm{Tr} e^{-\hat{H}_E}$.
In the case of free fermions, the entanglement Hamiltonian $\hat{H}_E$ should take a quadratic form as~\cite{Peschel2003}
\begin{equation}
\hat{H}_E=\sum_{ij} \hat{c}_i^{\dagger} (H_E)_{ij} \hat{c}_j =\sum_n \varepsilon_n \hat{a}_n^{\dagger} \hat{a}_n,
\end{equation}
where $\{\varepsilon_n\}$ are eigenvalues of the matrix $H_E$.
The single-particle entanglement spectrum $\{\xi_n\}$ can be defined as
\begin{equation}
\xi_n=\frac{1}{e^{\varepsilon_n}+1}.
\end{equation}
So the zero mode $\varepsilon_n=0$ for the entanglement Hamiltonian
corresponds to the mid-gap mode $\xi_n=0.5$ in the single-particle entanglement spectrum,
which indicates the nontrivial bulk topology for the many-body ground state~\cite{Fidkowski2010PRL}.
According to Ref.~\cite{Peschel2003}, the single-particle entanglement spectrum $\{\xi_n\}$ is given by
the eigenvalues of the correlation matrix $C_S$ for one subsystem $S$
\begin{equation}
[C_S]_{ij} = \mathrm{Tr} (\hat{\rho}_S \hat{c}_{j}^{\dagger} \hat{c}_{i}) = \langle \Psi_G | \hat{c}_{j}^{\dagger} \hat{c}_{i} | \Psi_G \rangle
\end{equation}
with $i,j$ denoting the degrees of freedom in $S$.
In fact, a real-space correlation matrix $C_S$ corresponds to a flat-band Hamiltonian in momentum space~\cite{Turner2010PRB,Hughes2011PRB}.

For 2D translational invariant insulators with the momenta $\bm{k}=(k_x,k_y)$ being good quantum numbers, e.g., Euler insulators that we study,
the many-body ground state can be represented as
\begin{equation}
|\Psi_G \rangle = \prod_{\bm{k},n\in \textrm{occ}} \hat{f}_{\bm{k},n}^{\dagger} |\mathrm{vac}\rangle
= \prod_{\bm{k},n\in \textrm{occ}} ( \sum_{\alpha} [u_{\bm{k}}^{n}]_{\alpha} \hat{c}_{\bm{k},\alpha}^{\dagger} ) |\mathrm{vac} \rangle ,
\end{equation}
where $\hat{f}_{\bm{k},n}^{\dagger}=\sum_{\alpha} [u_{\bm{k}}^{n}]_{\alpha} \hat{c}_{\bm{k},\alpha}^{\dagger}$
is the creation operator for the $n$th occupied eigenstate $|u_{\bm{k}}^{n}\rangle$ of the single-particle Hamiltonian $H(\bm{k})$.
Here $\alpha$ denotes the internal degrees of freedom within the unit cell.
In the following, we would like to consider two ways of real-space bipartition that cut along $x$ and $y$ directions, respectively.
For example, if we cut the system into two halves $X_1$ and $X_2$ along $y$ direction so that $k_y$ parallel to the cut remains a good quantum number.
Thus we have the correlation matrix $C_{X}(k_y)$ as a function of $k_y$ for the subsystem $X_1$ defined as
\begin{align}
[C_{X}(k_y)]_{x \alpha,x'\alpha'}
&= \left\langle \Psi_G|\hat{c}^{\dagger}_{x'\alpha',k_y} \hat{c}_{x\alpha, k_y}|\Psi_G \right\rangle \nonumber \\
&=\frac{1}{L_x} \sum_{k_x} e^{i k_x (x-x')} \sum_{n\in \textrm{occ}}
[u_{\bm{k}}^{n}]^{*}_{\alpha'} [u_{\bm{k}}^{n}]_{\alpha} \nonumber \\
&=\frac{1}{L_x} \sum_{k_x} e^{i k_x (x-x')} [P_{\textrm{occ}}(\bm{k})]_{\alpha \alpha'},
\end{align}
where $x,x'$ denote the unit cells in the subsystem $X_1$ with $L_x$ unit cells along the $x$ direction,
and $P_{\textrm{occ}}(\bm{k})=\sum_{n\in \textrm{occ}} |u_n(\bm{k})\rangle \langle u_n(\bm{k})|$ is the projector onto occupied bands.
By diagonalizing $C_{X}(k_y)$, we can obtain the momentum-resolved entanglement spectrum $\mathrm{ES}_x(k_y)$ as the set of eigenvalues $\{\xi_n(k_y)\}$.
The correlation matrix $C_{Y}(k_x)$ and the single-particle entanglement spectrum $\mathrm{ES}_y(k_x)$ can be defined similarly.

\begin{figure}[t]
\centering
\includegraphics[width=\linewidth]{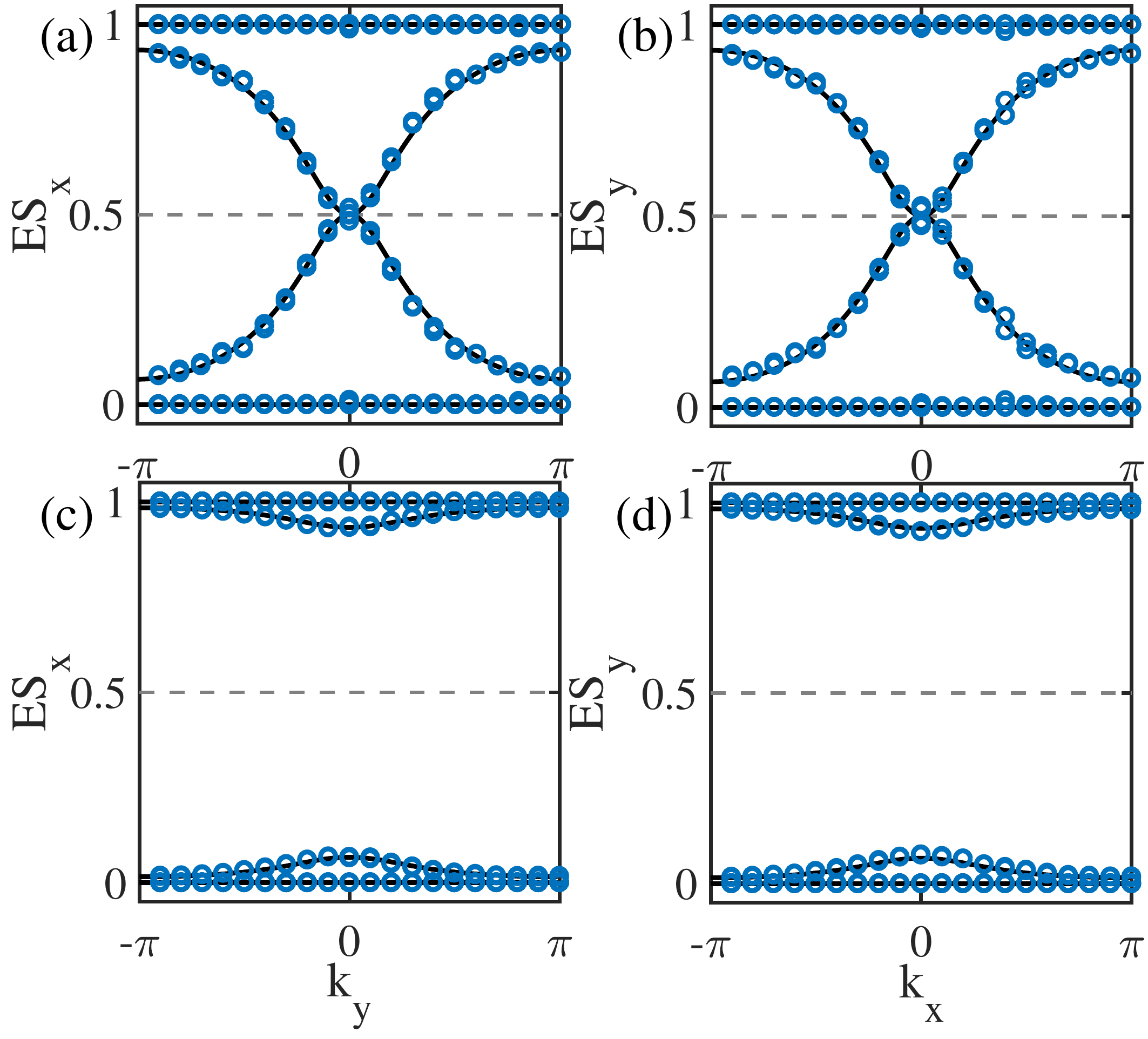}
\caption{\textbf{The measured entanglement spectra.}
The experimentally measured single-particle entanglement spectra $\mathrm{ES}_x(k_y)$ and $\mathrm{ES}_y(k_x)$
(denoted by circles) along with the theoretical values (denoted by solid lines) for the Euler Hamiltonian $H(\bm{k})$ with $20 \times 20$ unit cells.
In (a) and (b), $m=1$ corresponding to a nontrivial phase, and
in (c) and (d), $m=3$ corresponding to a trivial phase.
}
\label{fig_EntangleSpectra}
\end{figure}

Figure~\ref{fig_EntangleSpectra} displays the single-particle entanglement spectra $\mathrm{ES}_x(k_y)$ and $\mathrm{ES}_y(k_x)$
obtained from the experimentally measured unoccupied eigenstates $|u_3(\bm{k})\rangle$ for the three-band Euler Hamiltonian $H(\bm{k})$ with $20 \times 20$ unit cells.
In Fig.~\ref{fig_EntangleSpectra}(a),(b), we show the experimental and theoretical results
of the entanglement spectra $\mathrm{ES}_x$ and $\mathrm{ES}_y$ for the topologically nontrivial phase with $m=1$ in $H(\bm{k})$.
We can see that there exist mid-gap modes near $\xi_n=0.5$ for both of the entanglement spectra corresponding to the nontrivial Euler class, and
theoretical values show that the entanglement spectra have nonlinear dispersions near a four-fold degenerate point $\xi_n=0.5$,
which differs from the linear Dirac nodes in the single-particle entanglement spectra for inversion-symmetric topological insulators studied previously~\cite{Turner2010PRB,Hughes2011PRB}.
In Fig.~\ref{fig_EntangleSpectra}(c),(d), we show that
for the topologically trivial Euler insulator with $m=3$ in $H(\bm{k})$,
there is no mid-gap mode in the entanglement spectra $\mathrm{ES}_x(k_y)$ and $\mathrm{ES}_y(k_x)$
so that it corresponds to the trivial Euler class and can be adiabatically connected to a trivial phase with zero entanglement entropy.

\section*{Appendix H: The Hopf map in the quench dynamics of the Euler Hamiltonian}

\setcounter{equation}{0}
\renewcommand{\theequation}{H\arabic{equation}}
\renewcommand{\theHequation}{H\arabic{equation}}
\renewcommand{\bibnumfmt}[1]{[#1]} \renewcommand{\citenumfont}[1]{#1}

In this appendix, we will construct the Hopf map for the quench dynamics in the three-band Euler Hamiltonian
and show its relation with the quench dynamics in a two-band Chern insulator~\cite{Zhai2017PRL,Slager2020PRL}.

We first review the dynamical Hopf map for the quench dynamics in a two-band Chern insulator~\cite{Zhai2017PRL}.
We consider an initial state $ \psi_0(\bm{k},t=0) =\psi_0=(0,1)^T$ quenched by a two-band Chern Hamiltonian
$H_C(\bm{k})= \bm{d}(\bm{k}) \cdot\bm{\sigma}$ where $ \bm{d}(\bm{k}) =(d_x,d_y,d_z)^T$ is a unit vector at each momentum $\bm{k}=(k_x,k_y)$
in the 2D Brillouin zone. For the flattened postquench Hamiltonian,
the time-evolving state can be written as
\begin{align}
 \psi(\bm{k},t) &=e^{-it H_C(\bm{k})}  \psi_0 = [\cos(t) - i\sin(t) \bm{d} \cdot \bm{\sigma}] \psi_0 \nonumber \\
&=
\left(
  \begin{array}{c}
    -i\sin(t)(d_x-id_y) \\
    \cos(t)+i\sin(t)d_z \\
  \end{array}
\right).
\end{align}
Let us define a 4D vector $\bm{x}=(x_0,x_1,x_2,x_3)=(\cos(t),-\sin(t)d_x,-\sin(t)d_y,-\sin(t)d_z)$
which is a unit vector with $\sum_{i=0}^{3} x_i^2=1$ on $S^3$. Then the evolving state is
\begin{equation}\label{evolvingChernx}
 \psi(\bm{k},t) =
\left(
  \begin{array}{c}
    i x_1+x_2 \\
    x_0-i x_3 \\
  \end{array}
\right).
\end{equation}
Due to the periodicity of the evolving state in time $t$ with period $\pi$,
the space of $(k_x,k_y,t)$ forms a 3-torus $T^3$.
We can construct a map $f$ from $T^3$ to $S^2$ by mapping the evolving state onto a Bloch sphere $S^2$,
namely, for each $(k_x,k_y,t)$, the image $f(k_x,k_y,t)$ is the vector
$\hat{\bm{p}}= (p_x,p_y,p_z)=(\psi^{\dagger}  \sigma_x  \psi,  \psi^{\dagger}  \sigma_y  \psi, \psi^{\dagger}  \sigma_z  \psi)$ with
\begin{equation}
\sigma_x=
\begin{pmatrix}
  0 & 1 \\
  1 & 0 \\
\end{pmatrix},~~~
\sigma_y=
\begin{pmatrix}
  0 & -i \\
  i & 0 \\
\end{pmatrix},~~~
\sigma_z=
\begin{pmatrix}
  1 & 0 \\
  0 & -1 \\
\end{pmatrix}.
\end{equation}

The map $f$ from $T^3$ to $S^2$ can be decomposed into two steps as $f= h \circ g$~\cite{Deng2013PRB}.
The first map $g$ maps $(k_x,k_y,t)$ on $T^3$ to $(x_0,x_1,x_2,x_3)$ on $S^3$,
and the second map $h$ is a Hopf map in mathematics which maps $(x_0,x_1,x_2,x_3)$ on $S^3$ to $(p_x,p_y,p_z)$ on $S^2$.
The first map $g$ from $T^3$ to $S^3$ is classified  by an integer topological invariant~\cite{Neupert2012PRB}
\begin{equation}
\Gamma[g]=\frac{1}{12\pi^2}\int_{BZ\times [0,\pi]} \mathrm{d}^2 \bm{k} \mathrm{d}t ~ \epsilon_{\alpha\beta\gamma\rho}\epsilon_{\mu\nu\tau}
x_{\alpha}\partial_{\mu}x_{\beta}\partial_{\nu}x_{\gamma}\partial_{\tau}x_{\rho}.
\end{equation}
For $\bm{x}=(x_0,x_1,x_2,x_3)=(\cos(t),-\sin(t)d_x,-\sin(t)d_y,-\sin(t)d_z)$, a straightforward calculation yields
\begin{equation}
\Gamma[g]=-\frac{1}{4\pi} \int_{BZ}\mathrm{d}^2 \bm{k} [\bm{d} \cdot (\partial_{k_x}\bm{d}\times \partial_{k_y}\bm{d})],
\end{equation}
which is equivalent to the winding number of $\bm{d}(\bm{k})$ over the 2D Brillouin zone.
The second map is a Hopf map that maps $(x_0,x_1,x_2,x_3)$ on $S^3$ to $\hat{\bm{p}}$ on $S^2$ as
\begin{equation}\label{Hopf1}
\hat{\bm{p}}=
\left(
  \begin{array}{c}
    p_x \\
    p_y \\
    p_z \\
  \end{array}
\right)=
\left(
  \begin{array}{c}
    2 x_0 x_2-2 x_1 x_3 \\
    -2 x_0 x_1-2 x_2 x_3 \\
    x_1^2+x_2^2-x_0^2-x_3^2 \\
  \end{array}
\right).
\end{equation}
This Hopf map is nontrivial and characterized by a unit Hopf index~\cite{Moore2008PRL,Deng2013PRB}.
The topological property of the map $f$ is determined by the product of the topological invariants of the two maps $g$ and $h$~\cite{Deng2013PRB}.
Therefore, this map is nontrivial for a nontrivial postquench Chern Hamiltonian with nonzero winding number of $\bm{d}(\bm{k})$ (Chern number),
which leads to the corresponding linking structure for the inverse images of the map in the quench dynamics~\cite{Zhai2017PRL}.

Now we turn to discuss the quench dynamics for the three-band flattened Euler Hamiltonian $H_E(\bm{k})=2\bm{n}(\bm{k})\cdot\bm{n}(\bm{k})^{T}-I$.
Starting from the initial state $ \psi_0(\bm{k},t=0)=\psi_0 =(0,0,1)^T$, we have the time-evolving state
\begin{align}
 \psi(\bm{k},t) &=e^{-itH_E(\bm{k})} \psi_0 = [\cos(t)-i\sin(t)H_E] \psi_0 \nonumber \\
 &= \cos(t)\psi_0 -i \sin(t) \bm{a}(\bm{k}) \nonumber \\
&=
\left(
  \begin{array}{c}
    -i\sin(t)a_x \\
    -i\sin(t)a_y \\
    \cos(t)-i\sin(t)a_z \\
  \end{array}
\right),
\end{align}
where we define $\bm{a}({k})=(a_x,a_y,a_z)^T=H_E(\bm{k})\psi_0$ which is also a real unit vector due to the reality of $H_E(\bm{k})$.
We can define a 4D unit vector on $S^3$ as $\bm{x}=(x_0,x_1,x_2,x_3)=(\cos(t),\sin(t)a_x,\sin(t)a_y,\sin(t)a_z)$,
which is similar to the vector $\bm{x}$ for quenching the Chern Hamiltonian by replacing $-\bm{d}(\bm{k})$ to $\bm{a}(\bm{k})$.
Then the evolving state is
\begin{equation}\label{evolvingEulerx}
 \psi(\bm{k},t) =
\left(
  \begin{array}{c}
    -i x_1 \\
    -i x_2 \\
    x_0-i x_3 \\
  \end{array}
\right),
\end{equation}
which also has a period $\pi$ in time.

Analogously to the quench dynamics of a Chern Hamiltonian, we can construct a similar map $f$ from $T^3$ to $S^2$ as
the composition of two maps $g$ and $h$.
The first map $g$ from $T^3$ to $S^3$ is the same as the Chern Hamiltonian case with $\bm{d}(\bm{k})$ replaced by $-\bm{a}(\bm{k})$.
For the Hopf map $h$, we define the image vector $\hat{\bm{p}}$ on $S^2$ for a 4D vector $\bm{x}=(x_0,x_1,x_2,x_3)$
as $\hat{\bm{p}}= (p_x,p_y,p_z)=(\psi^{\dagger}  \mu_x  \psi,  \psi^{\dagger}  \mu_y  \psi, \psi^{\dagger}  \mu_z  \psi)$,
where the three components are the expectation values of three Hermitian operators as follows
\begin{align}
&\mu_x=
\begin{pmatrix}
  0 & 0 & -1 \\
  0 & 0 & -i \\
  -1 & i & 0 \\
\end{pmatrix},~~~
\mu_y=
\begin{pmatrix}
  0 & 0 & i \\
  0 & 0 & -1 \\
  -i & -1 & 0 \\
\end{pmatrix},~~~ \nonumber \\
&\mu_z=
\begin{pmatrix}
  1 & 0 & 0 \\
  0 & 1 & 0 \\
  0 & 0 & -1 \\
\end{pmatrix}.
\end{align}
In fact, these three Hermitian matrices can be constructed from the Pauli matrices for the quench dynamics of the Chern Hamiltonian by noticing the relation that
we can obtain the evolving state (\ref{evolvingChernx}) for the Chern Hamiltonian
after linearly combining the first two components in the evolving state (\ref{evolvingEulerx}) for the Euler Hamiltonian into a single component.
One can easily verify that the map defined above is the same as the Hopf map defined for the Chern Hamiltonian (\ref{Hopf1}).

We therefore establish the correspondence between the quench dynamics in a three-band Euler Hamiltonian and the dynamics in
a two-band Chern Hamiltonian by identifying the similar roles of $\bm{a}(\bm{k})$ and $\bm{d}(\bm{k})$ in the Hopf maps.
The fact that the nontrivial winding of $\bm{d}(\bm{k})$ over the Brillouin zone gives rise to the nontrivial Hopf links in the quench dynamics of the Chern Hamiltonian suggests that
in the quench dynamics of the Euler Hamiltonian,
if the vectors $\bm{a}(\bm{k})$ wrap the sphere $S^2$ for nonzero times in a closed region of the Brillouin zone,
there should exist nontrivial Hopf links in the corresponding region of momentum-time space.
As shown in the main text, for the topologically nontrivial postquench Euler Hamiltonian with $\xi=2$,
$\bm{a}(\bm{k})$ has opposite winding numbers $\pm 1$ in two separate patches of Brillouin zone
so that we have a pair of Hopf link and antilink with opposite linking number given by the Chern number for the Hamiltonian $H_C(\bm{k})=\bm{a}(\bm{k})\cdot \bm{\sigma}$
in the two regions, respectively.
The linking number of the Hopf links can be determined by calculating the Hopf invariant~\cite{Wilczek1983PRL,Zhai2017PRL}
\begin{equation}
\chi=\int_{BZ \times [0,\pi]} \mathrm{d}^2 \bm{k} \mathrm{d}t B_{\mu} J^{\mu},
\end{equation}
where $\bm{J}$ and $\bm{B}$ are determined by
\begin{align}
&J^{\mu}=\frac{1}{8\pi} \epsilon^{\mu \nu \lambda} \hat{\bm{p}} \cdot (\partial_{\nu} \hat{\bm{p}} \times \partial_{\lambda} \hat{\bm{p}}) \\
&J^{\mu}=\epsilon^{\mu \nu \lambda} \partial_{\nu} B_{\lambda}
\end{align}
with $\mu=k_x,k_y,t$ and $\hat{\bm{p}}=\hat{\bm{p}}(k_x,k_y,t)$ being the image of the map $f$ from the $(k_x,k_y,t)$ space to $S^2$.
For quench dynamics of the Euler Hamiltonian discussed above,
our numerical calculations indeed show that $\chi=\pm 1$ for the map $f$ restricted on the two separate regions of the momentum-time space with opposite winding of $\bm{a}(\bm{k})$ in the corresponding patches of 2D Brillouin zone, respectively.
Therefore, there exists a pair of Hopf link and antilink with opposite linking numbers in momentum space, which are closely related to
the skyrmion and antiskyrmion structure for $\bm{a}(\bm{k})$ in the Brillouin zone.

\end{document}